\documentclass{article}

\usepackage{arxiv}

\usepackage[utf8]{inputenc} 
\usepackage[T1]{fontenc}    
\usepackage{hyperref}       
\usepackage{url}            
\usepackage{booktabs}       
\usepackage{amsfonts}       
\usepackage{nicefrac}       
\usepackage{microtype}      
\usepackage{lipsum}
\usepackage[justification=centering]{caption}
\usepackage[bottom]{footmisc}
\usepackage{amsmath}
\usepackage{graphicx}
\usepackage{float}
\usepackage{array}
\usepackage{subcaption,graphicx}
\usepackage{setspace}
\usepackage{lscape} 
\usepackage{mathtools}
\usepackage{breakcites}
\usepackage{indentfirst}
\usepackage{indentfirst}
\usepackage{mwe}
\usepackage{float}
\usepackage{booktabs}
\usepackage{url}
\usepackage{mathrsfs}

\title{Power Analysis for Stepped Wedge Trials with Two Treatments}
\author{
Phillip T Sundin\thanks{Corresponding author}\\
  Department of Biostatistics\\
  University of California, Los Angeles\\
  Los Angeles, CA 90049 \\
  \texttt{phillip1492@ucla.edu} \\
   \And
 Catherine M Crespi \\
  Department of Biostatistics\\
  University of California, Los Angeles\\
  Los Angeles, CA 90049 \\
  \texttt{ccrespi@ucla.edu} \\
}

\begin{document}
\maketitle

\begin{abstract}
Stepped wedge designs (SWDs) are designs for cluster randomized trials that feature staggered, unidirectional cross-over, typically from a control to a treatment condition. Existing literature on statistical power for SWDs primarily focuses on designs with a single treatment. However, SWDs with multiple treatments are being proposed and conducted. We present a linear mixed model for a SWD with two treatments, with and without an interaction between them. We derive closed form solutions for the standard errors of the treatment effect coefficients for such models along with power calculation methods. We consider repeated cross-sectional designs as well as open and closed cohort designs and different random effect structures. Design features are examined to determine their impact on power for main treatment and interaction effects.
\end{abstract}


\section{Introduction}   
\begin{figure}[ht]
	\centering
	\includegraphics[scale=0.35]{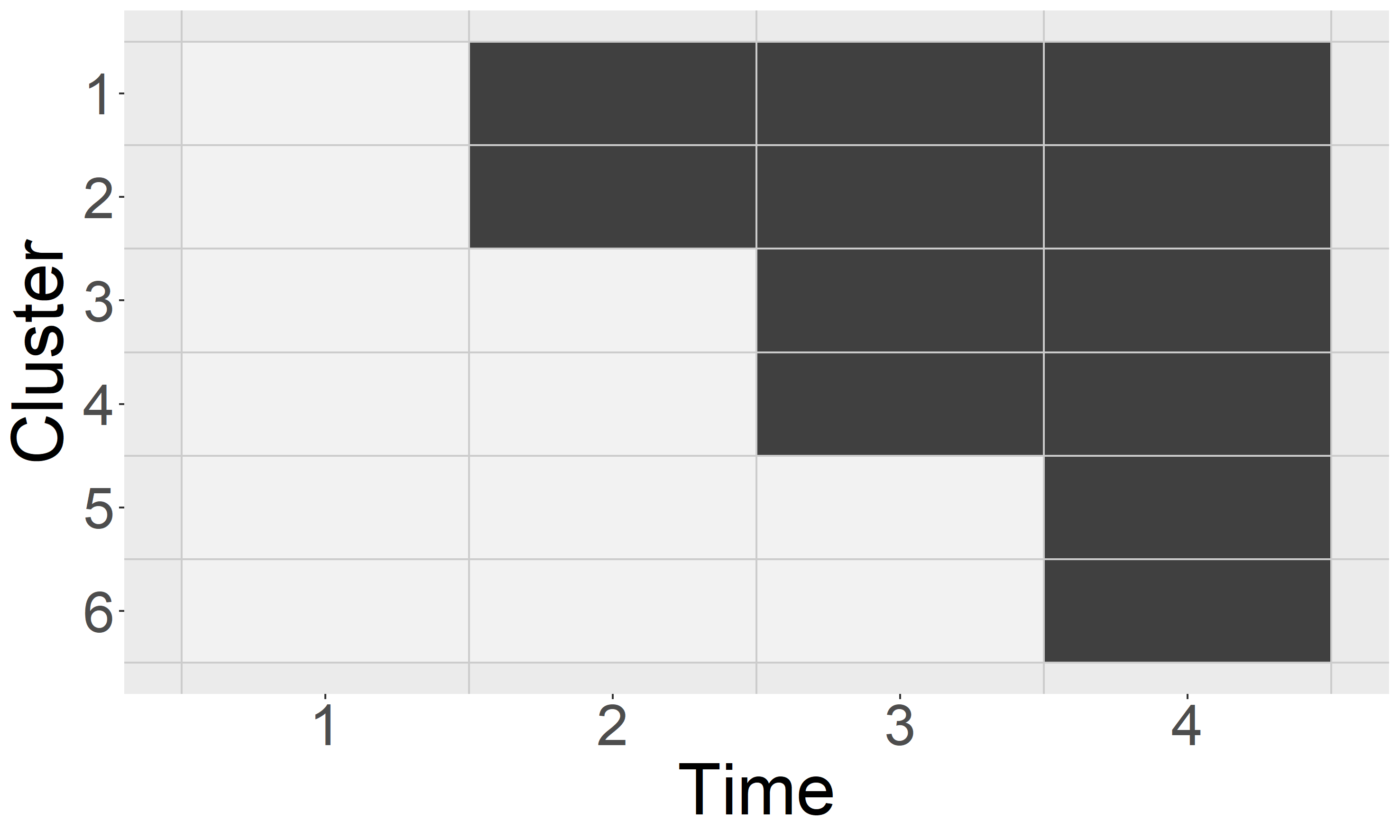}
	\caption{Example of a stepped wedge design with six clusters, four time periods, and one treatment. There are three distinct sequences with two clusters in each sequence. A white cell indicates the control condition and a gray cell represents the treatment condition.}
	\label{Figure:Example}

\end{figure}

\doublespacing
Cluster randomized trials (CRTs) are trials in which groups of individuals, called clusters, are randomized to treatment arms. CRTs can use parallel designs, in which clusters receive only one treatment, or crossover designs, in which clusters are assigned to a sequence of treatments. A variation on the CRT crossover design is the stepped wedge design (SWD) \cite{HusseyHughes}. The most common SWDs involve unidirectional crossover in which clusters transition from a control condition to a treatment condition at different pre-specified times. Clusters are randomized to the pre-determined sequences. Typically, all clusters are in the treatment condition in the last period \cite{AnalysisCRTRepeatedCS}. An example of such a SWD is shown in Figure \ref{Figure:Example}.

SWDs have several advantages over parallel and crossover CRT designs. SWDs allow comparisons both within cluster and across cluster, which can yield efficiency gains \cite{ReduceSampleSize}. It may be less costly and logistically easier to roll out the intervention over time instead of all at once, as would occur in many parallel CRTs \cite{ReviewReportingQualityDesignFeatures}. Guaranteeing treatment for all clusters may make clusters more willing to participate or alleviate some ethical concerns \cite{ReduceSampleSize}.
	
Most research on the design and analysis of stepped wedge trials has focused on the SWD with one intervention contrasted with a control condition. There is a small but growing body of literature on SWDs involving multiple interventions. Grayling et al. \cite{Admissable_Multiarm} focus on studies in which there is a nested natural order of $D$ interventions such that intervention $d$ consists of intervention $d-1$ plus some additional factor. The authors discuss the optimization of treatment sequence allocations and focus on optimal design for such trials. Variances for treatment effect estimates in SWDs with nested interventions have also been studied \cite{Multiarm_Variance}. The limited existing literature does not include modeling of a potential interaction between treatments \cite{ProposedVariations}. 

SWDs with multiple interventions are being implemented in practice despite a lack of methodological literature. For example, a recently funded trial of the comparative effectiveness of two interventions to promote human papillomavirus (HPV) vaccination is using a stepped wedge design implemented at seven clinics over a four-year period \cite{Study_Bastani}. One intervention is a parent reminder sent via text message, and the other is a multi-component clinic-based intervention comprised of staff education, workflow modifications, and regular auditing. The study combines the stepped wedge design with a $2\times2$ factorial design, with cluster-periods in a control condition, single intervention conditions for both treatments, and a combined condition  in which clinics receive both interventions. In another example, the STARSHIP study examined two treatments to assess their efficacy in reducing hyperbilirubinaemia in infants \cite{SWD_FactorialExample}. The study employed a SWD with two treatments and a combined condition where a clinic receives both treatments simultaneously. The STARSHIP study assumed additive treatment effects and did not consider an interaction effect.  Lyons et al.$\;$identified additional studies that employed SWDs with more than one treatment \cite{ProposedVariations}. One such example is the SSTP-ACT study that compared two behavioral health interventions with the goal of reducing behavioral problems in children with cerebral palsy \cite{BehavioralProblems}. Similar to the STARSHIP study, the SSTP-ACT study had a combined condition where clusters can receive both treatments simultaneously but did not include an interaction effect.  

There are also examples in the literature of studies that conduct two one-treatment SWD trials simultaneously. The FallDem study examined two interventions for improving the lives of dementia patients \cite{ProposedVariations, Dementia}. Durovni et al.\@ conducted two SWD trials for tuberculosis screening \cite{TBScreening, Durovni_Tuberculosis2}. It is unclear whether the investigators of these studies considered combining the two separate SWDs into one larger SWD with two treatments. This might have allowed for more precise estimates of treatment effects, direct comparisons of the treatments, and a possible reduction in the number of clusters required to achieve the desired level of power.

In this paper, we consider SWDs with two treatments, with and without a combined condition, and with and without an interaction effect. We call trials that include a combined condition stepped wedge factorial designs (SWFDs). As in other $2\times2$ factorial designs, these trials involve two factors, each with two levels, with four treatment combinations in total. Factorial designs have several well-known advantages, such as potentially increased efficiency and the ability to study interaction effects \cite{FirstcourseDOE}.  We derive analytical results for conducting power analysis for such trials and examine factors that influence power. 

The paper is organized as follows. Section \ref{Section:Methods} introduces the models for the SWD with two treatments including the SWFD model and develops power analysis methods. Section \ref{Section:DesignIssues} uses illustrative examples to examine the influence of different design features on power. Section \ref{Section:Discussion} discusses our findings, possible extensions, limitations and future work.

\section{Methods}
\setlength{\abovedisplayskip}{10pt}
\setlength{\belowdisplayskip}{10pt}
\setlength{\abovedisplayshortskip}{10pt}
\setlength{\belowdisplayshortskip}{10pt}

\label{Section:Methods}
We first present models for SWDs with two treatments, with and without an interaction term, assuming a repeated cross-sectional study design in which outcomes are assessed on different individuals within each cluster in each time period. We then discuss cohort designs, in which outcomes are measured on some or all of the same individuals in a cluster across time periods, and the nested exchangeable model, which relaxes some assumptions. We standardize the models to allow power calculations to be conducted by specifying intracluster correlation (ICC) and autocorrelation values instead of variance components, which can be difficult to specify. Power for both main effects and an interaction effect are considered. The section concludes with an overview of the derivation of closed form solutions for the standard errors of the estimated treatment effect coefficients, with the full derivation found in Appendix A. 

\subsection{Cross-Sectional Design}
\label{Subsection:CrossSectional}
We begin with a model for a single binary treatment factor and a repeated cross-sectional design, as originally proposed by Hussey and Hughes \cite{HusseyHughes}. For a design with $I$ clusters observed at $T$ times, and $N$ different individuals per time per cluster, let $Y_{ijk}$ be a continuous outcome for for individual $k$ in cluster $i$ at time $j$. The model for $Y_{ijk}$ is
	\begin{equation}
	Y_{ijk} = \mu + \alpha_i + \beta_j + X_{ij}\theta_1 + e_{ijk}
	\label{Equation:HusseyHughes}
	\end{equation}
where $\mu$ is an intercept, $\alpha_i \sim N(0,\sigma_{\alpha}^2)$ is a random intercept for cluster $i$, $\beta_j$ is a fixed effect for time $j$, $X_{ij}$ is a \{0,1\} indicator for whether cluster $i$ at time $j$ receives treatment, $\theta_1$ is the treatment effect, and $e_{ijk} \sim N(0,\sigma_{e}^2)$.  The total variance of an individual level outcome is $\sigma_y^2=\sigma_{\alpha}^2 + \sigma_{e}^2$.

This model can easily be expanded to accommodate multiple binary treatment factors. Assuming additive treatment effects, the model with two treatment conditions is
	\begin{equation}
	\label{Equation:NoInteractionModel}
	{Y}_{ijk} = \mu + \alpha_i + \beta_j + X_{ij}\theta_1 + W_{ij}\theta_2 + e_{ijk}, 
	\end{equation} where $W_{ij}$ is a \{0,1\} indicator of whether cluster $i$ at time $j$ receives treatment 2 and $\theta_2$ is the treatment effect for treatment 2.  Adding an interaction effect to accommodate trials  in which a cluster can receive both treatments simultaneously, the model becomes
	\begin{equation}
	\label{Equation:MainSWFDModel_Individual}
	Y_{ijk} = \mu + \alpha_i + \beta_j + X_{ij}\theta_1 + W_{ij}\theta_2 + X_{ij}W_{ij}\theta_{3} + e_{ijk},\end{equation}
 where $\theta_{3}$ is the interaction effect.
 
 In these models, the correlation between outcomes of two different individuals $k$ and $k'$ in the same cluster $i$ at the same time $j$, Corr($y_{ijk},y_{ijk'}$), is commonly termed the intracluster correlation (ICC). We denote this ICC as $\rho_w$. It can be shown that $\rho_w = \sigma_{\alpha}^2/\sigma_{Y}^2$. These models also imply that the correlation between two observations on different individuals in the same cluster and same time period is equal to the correlation between two observations in the same cluster but different time periods, i.e., Corr($y_{ijk},y_{ijk'}$) = Corr($y_{ijk},y_{ij'k'}$). We denote the latter correlation as $\rho_a$ and note that the repeated cross-sectional model assumes $\rho_w=\rho_a$. 
 
 To standardize one of these models, one can divide it through by $\sigma_y$. The cluster random intercept $\alpha_i$ now has standardized variance $\rho_w$ and the error term has standardized variance $1 - \rho_w$. Standardization is convenient for power calculations because only $\rho_w$ needs to be specified rather than both $\sigma_e^2$ and  $\sigma_\alpha^2$.

To derive standard errors, it is convenient to work with cluster-level outcomes. Let $\overline{Y}_{ij\cdot} = \frac{1}{N}\sum_{k=1}^{N}Y_{ijk}$ be the mean outcome of cluster $i$ at time $j$ across $N$ individuals. The model for cluster-period means is
	\begin{equation}
	\overline{Y}_{ij\cdot} = \mu + \alpha_i + \beta_j + X_{ij}\theta_1 + W_{ij}\theta_2 + X_{ij}W_{ij}\theta_{3}+ e_{ij\cdot} \label{Equation:MainSWDModel},
	\end{equation}
where $e_{ij\cdot}  = \frac{1}{N}\sum_{k=1}^{N}e_{ijk} \sim N(0,\sigma_{c}^2)$ and $\sigma_{c}^2 = \frac{\sigma_{e}^2}{N}.$ Under this model, the variance of a cluster-period mean is $\text{Var}(\overline{Y}_{ij\cdot}) = \sigma_{c}^2 + \sigma_{\alpha}^2$. Define the outcome vector $\boldsymbol{Y} = (\overline{Y}_{11\cdot},\dots,\overline{Y}_{iT\cdot},\dots,\overline{Y}_{I1\cdot},\dots,\overline{Y}_{IT\cdot})' $. Assuming clusters are independent, the variance-covariance matrix of $\boldsymbol{Y}$ is a $IT\times IT$ matrix of the form
	\[\boldsymbol{V}=
	\begin{bmatrix}
	\boldsymbol{V_1} & 0 & 0  & 0 \\
	0 & \boldsymbol{V_2} & \dots & 0 \\
	0 & 0 & \ddots & 0 \\
	0 & 0 & \dots & \boldsymbol{V_I} \\
	\end{bmatrix}, 
	\] 
	with each $T\times T$ matrix $\boldsymbol{V_i}$ having structure	\[\boldsymbol{V_i} =
	\begin{bmatrix}
	\sigma_{\alpha}^2 +\sigma_{c}^2 & \sigma_{\alpha}^2& \dots  & \sigma_{\alpha}^2 \\
	\sigma_{\alpha}^2 & \sigma_{\alpha}^2+\sigma_{c}^2 & \dots &  \sigma_{\alpha}^2 \\
	\vdots & \vdots  & \ddots & \vdots \\
\sigma_{\alpha}^2&\sigma_{\alpha}^2 & \dots &\sigma_{\alpha}^2+\sigma_{c}^2 \\
	\end{bmatrix}.
	\]

In the standardized model, $\boldsymbol{V_i}$ will have diagonal elements $\rho_w + (1-\rho_w)/N$ and off-diagonal elements $\rho_w$. Now we turn to the design matrix of the fixed effects. Setting $\beta_T= 0$ for identifiability \cite{HusseyHughes}, the $(T+3)\times1$ regression coefficient vector for the fixed effects is
	\[ \boldsymbol{\eta} = 
	\begin{bmatrix}
	\mu &
	\beta_1 &
	\ldots &
	\beta_{T-1} &
	\theta_1 & \theta_2 & \theta_{3}
	\end{bmatrix}^{'}.
	\] 	
Then, the full $IT \times(T+3)$ design matrix $\boldsymbol{Z}$ becomes 
\[\boldsymbol{Z} = 
\begin{bmatrix}
	\boldsymbol{Z_1} &
	\boldsymbol{Z_2} &
	\ldots &
	\boldsymbol{Z_I}
	\end{bmatrix}^{'}. \] 
Each matrix $\boldsymbol{Z_i}$ has dimension $\text{T}\times\text{(T+3)}$  and takes the form
\[\boldsymbol{Z_i} = 
	\begin{bmatrix}
	\boldsymbol{1_T} &
	\begin{matrix}
	\boldsymbol{I_{T-1}} \\
	\boldsymbol{0^{'}_{T-1}}
	\end{matrix} &
	\boldsymbol{X_i} & \boldsymbol{W_i} & \boldsymbol{(XW)_i}
	\end{bmatrix}\text{.}
	\] The elements of the vector $\boldsymbol{X_i} = (X_{i1},X_{i2},\dots,X_{iT})'$ are indicators of whether cluster $i$ at time $j$ receives treatment 1, the elements of $\boldsymbol{W_i} = (W_{i1},W_{i2},\dots,W_{iT})'$ are indicators of receipt of treatment 2, and $\boldsymbol{(XW)_i}$ is the Hadamard product of $\boldsymbol{X_i} \text{ and } \boldsymbol{W_i}$, with a value of 1 if cluster $i$ receives both treatments at time $j$ and 0 otherwise. The matrix $\boldsymbol{I_{T-1}}$ contains indicators for each time $j$ from $1,\dots,(T-1)$. The vector $\boldsymbol{0^{'}_{T-1}}$ corresponds to time $T$.
	
\subsection{Cohort Design}
\label{Subsection:ClosedCohort}
When an individual can be observed at more than one time period, we must account for the correlation between observations from the same individual. This can be accomplished by adding a random intercept $\psi_{ik} \sim N(0,\sigma_{\psi}^2)$ for individual $k$ in cluster $i$ to the cross-sectional model (\ref{Equation:MainSWFDModel_Individual}), yielding 	\begin{equation}
	Y_{ijk} = \mu + \alpha_i + \beta_j + X_{ij}\theta_1 + W_{ij}\theta_2 + X_{ij}W_{ij}\theta_{3} + \psi_{ik} + e_{ijk}.
	\label{Equation:IndividualSWDModelLongitudinal}
	\end{equation}
The total variance of an individual level outcome in this model is $\sigma_y^2=\sigma_{\alpha}^2+\sigma_{\psi}^2+\sigma_{e}^2$. The correlation between two observations in the same cluster and same time period, Corr($y_{ijk},y_{ijk'}$), $k\neq k'$, is $\rho_w = \sigma_{\alpha}^2 /\sigma_{Y}^2$, and the correlation between two observations on the same individual $k$ at times $j$ and $j`$, Corr($y_{ij'k},y_{ijk}$), $j\neq j'$, is $\rho_a = (\sigma_{\alpha}^2 + \sigma_{\psi}^2)/\sigma_{Y}^2$. Note that $\rho_a \geq \rho_w$.

Individual auto-correlation (IAC) is defined as the proportion of the individual-level variance (which in this model is $\sigma_{\psi}^2 + \sigma_{e}^2$) that is time-invariant. In this model, the IAC is $\pi = \sigma_{\psi}^2/(\sigma_{\psi}^2 + \sigma_{e}^2)$. Setting $\pi = 0$ yields a repeated cross-sectional design. By allowing $\pi$ to vary between 0 and a specified maximum value for individuals followed throughout the entire study, this model can handle studies with a mixture of cross-sectional and cohort observations for power calculations \cite{ANCOVA_CRT,CohortCrossSectionalDesign}, including open cohort designs in which individuals can enter and/or depart during the study period \cite{SWDTerminology}.

Dividing the model by $\sigma_{Y}$, we obtain the standardized variance of $\alpha_i$ as $\rho_w = \sigma^2_\alpha/\sigma_{Y}^2$. Using the definition of the IAC and the fact that the standardized variances sum to 1, we obtain the standardized variance of $\psi_{ik}$ as $\pi(1 - \rho_w)$ and the standardized variance of the error term as $1 - \rho_w  - \pi(1 - \rho_w)$. Thus we can specify all standardized variances in terms of $\rho_w$ and $\pi$, which is convenient for power calculations.

In this model, the cluster-period means are equal to
	\begin{equation}
	\overline{Y}_{ij\cdot} = \mu + \alpha_i + \beta_j + X_{ij}\theta_1 + W_{ij}\theta_2 + X_{ij}W_{ij}\theta_{3} + \psi_{i} + e_{ij}, \label{Equation:SWDModelLongitudinal}
	\end{equation}
	where $\psi_{i} = \frac{1}{N}\sum_{k=1}^{N}\psi_{ik} \sim N(0, \frac{\sigma_{\psi}^2}{N})$.  The variance-covariance matrix of the cluster-period means for cluster $i$ is 
	
	\[ \hspace{-0.5cm} \boldsymbol{V_i} =
	\begin{bmatrix}
	\sigma_{\alpha}^2+\sigma_{c}^2 + \frac{\sigma_{\psi}^2}{N} & \sigma_{\alpha}^2  + \frac{\sigma_{\psi}^2}{N} & \dots  &\sigma_{\alpha}^2+ \frac{\sigma_{\psi}^2}{N} \\
	\sigma_{\alpha}^2  + \frac{\sigma_{\psi}^2}{N} & \sigma_{\alpha}^2+\sigma_{c}^2  + \frac{\sigma_{\psi}^2}{N} & \dots &  \sigma_{\alpha}^2 + \frac{\sigma_{\psi}^2}{N} \\
	\vdots & \vdots  & \ddots & \vdots \\
	\sigma_{\alpha}^2 + \frac{\sigma_{\psi}^2}{N} & \sigma_{\alpha}^2 + \frac{\sigma_{\psi}^2}{N}& \dots & \sigma_{\alpha}^2+\sigma_{c}^2  + \frac{\sigma_{\psi}^2}{N} \\
	\end{bmatrix}.
	\] In the standardized model, the matrix $\boldsymbol{V_i}$ has diagonal elements $\rho_w + (1 - \rho_w)/N$ and off-diagonal elements $ \rho_w + \pi(1-\rho_w)/N$.
	
\subsection{Nested Exchangeable Correlation Model}
\label{Subsection:TimeTrend}
The repeated cross-sectional model (\ref{Equation:MainSWFDModel_Individual}) assumes the correlation between two observations in the same cluster and same time period is equal to the correlation between two observations in the same cluster but different time periods, i.e., $\rho_w=\rho_a$. Using the approach in Hemming et al. \cite{AnalysisCRTRepeatedCS}, one can allow these value to differ by adding a random effect $\nu_{ij}$ for cluster $i$ in time $j$, yielding the model \begin{equation}
	Y_{ijk} = \mu + \alpha_i + \beta_j + X_{ij}\theta_1 + W_{ij}\theta_2 + X_{ij}W_{ij}\theta_{3} + \nu_{ij} + e_{ijk}, \;\;\;  
	\label{Equation:VaryingSecularIndividual}
	\end{equation}
in which $\nu_{ij}\sim N(0,\sigma_{\nu}^2)$ is assumed independent of $e_{ijk}$ and $\alpha_i$. In this model, $\sigma_y^2 = \sigma_{\nu}^2 + \sigma_{\alpha}^2 + \sigma_{e}^2$. This approach divides the cluster-level variance into time-varying and time invariant components, namely, $\sigma_{\nu}^2$ and $\sigma_{\alpha}^2$.  The within-period ICC is now $\rho_w = (\sigma_{\nu}^2 + \sigma_{\alpha}^2)/ \sigma_y^2$ and the across-period ICC is $\rho_a = \sigma_{\alpha}^2 / \sigma_y^2 .$ 
Let the cluster auto-correlation (CAC) be the proportion of cluster level variance that is time-invariant \cite{ANCOVA_CRT, CohortCrossSectionalDesign}. Then we have CAC $ = \sigma_{\alpha}^2/(\sigma^2_{\nu} + \sigma_{\alpha}^2) = \rho_a/\rho_w.$ 

The standardized variances of $\alpha_i$, $\nu_{ij}$ and $e_{ijk}$ are $\rho_a$, $\rho_w - \rho_a$ and $1-\rho_w$, respectively. Thus all variance components can be expressed in terms of $\rho_w$ and $\rho_a$. Note that we must have $\rho_w \geq \rho_a$.

Under this model, cluster-period means take the form
\begin{equation}
	\overline{Y}_{ij\cdot} = \mu + \alpha_i + \beta_j + X_{ij}\theta_1 + W_{ij}\theta_2 + X_{ij}W_{ij}\theta_{3} + \nu_{ij} + e_{ij\cdot}
	\label{Equation:VaryingSecular}
	\end{equation} and the variance-covariance matrix for cluster $i$ is
	\[ \hspace{-0.5cm} \boldsymbol{V_i} =
	\begin{bmatrix}
	\sigma_{\alpha}^2 + \sigma_{c}^2 + \sigma_{\nu}^2& \sigma_{\alpha}^2  & \dots  &\sigma_{\alpha}^2 \\
	\sigma_{\alpha}^2   & \sigma_{\alpha}^2+\sigma_{c}^2+ \sigma_{\nu}^2 & \dots &  \sigma_{\alpha}^2  \\
	\vdots & \vdots  & \ddots & \vdots \\
	\sigma_{\alpha}^2  & \sigma_{\alpha}^2& \dots & \sigma_{\alpha}^2+\sigma_{c}^2 + \sigma_{\nu}^2\\
	\end{bmatrix}.
	\] In the standardized model, $\boldsymbol{V_i}$ has diagonal elements $\rho_w + (1 - \rho_w)/N$ and off-diagonal elements  $\rho_a$.


\subsection{Power Analysis}

Inference for fixed effects in linear mixed models can be conducted using Wald tests or likelihood ratio tests. We focus on Wald tests. For hypotheses of the form $H_0: \eta = 0$, where $\eta$ is a fixed effects coefficient, the Wald test statistic takes the form $\hat{\eta}/\sqrt{\text{Var}(\hat{\eta}})$, where $\hat{\eta}$ is the estimated coefficient, and has an approximate standard normal distribution when the null is true \cite{MixedEffectsBook2}. The power to reject $H_0$ for a specific true value of $\eta$, denoted $\eta_a$, with type I error rate $\alpha$ is approximately $$ P\Bigg( \; \Biggl|{\frac{\eta_a}{\sqrt{\text{Var}(\hat{\eta})}}}\Biggl|\;\; \geq \;\; z_{1-\frac{\alpha}{2}} \; \Big| \; \eta = \eta_a
\Bigg)$$
where $z_{1-\frac{\alpha}{2}}$ is the corresponding $(1-\frac{\alpha}{2})^{th}$ percentile of the standard normal distribution.

To calculate power, we need an expression for $\text{Var}(\hat{\eta})$. We derive expressions for $\text{Var}(\hat{\eta})$ for each of the three models using the cluster-period mean models in (\ref{Equation:MainSWDModel}), (\ref{Equation:SWDModelLongitudinal}) and (\ref{Equation:VaryingSecular}). Given that these are linear mixed models, the variance-covariance matrix of the estimated fixed effect coefficients takes the form $\boldsymbol{C} = (\boldsymbol{Z^{'}V^{-1}Z})^{-1}$, where $\boldsymbol{Z}$ is the fixed effects design matrix and  $\boldsymbol{V}$ is the variance-covariance matrix of the outcome vector. We are most interested in finding expressions for the variances and covariances of treatment effect coefficient estimates, $\hat\theta_1$, $\hat\theta_2$, and $\hat\theta_{3}$. We find an expression for $\boldsymbol{Z^{'}V}^{-1}\boldsymbol{Z}$ then invert it to get the elements of $(\boldsymbol{Z^{'}V}^{-1}\boldsymbol{Z})^{-1}$ corresponding to these variances and covariances.  
We focus on the repeated cross-sectional model here; the derivation generalizes to the cohort and nested exchangeable models by using the appropriate structure of the matrix $\boldsymbol{V}$. 

Let $\boldsymbol{Z}$ be the $IT \times (T+3)$ design matrix and $\boldsymbol{V}$ be the $IT \times IT$ variance-covariance matrix of the cluster-level outcomes. Assuming clusters are independent, the matrix $\boldsymbol{V}$ has block diagonal structure with elements $\boldsymbol{V_i} = \sigma_c^2\boldsymbol{I_T} + \sigma_{\alpha}^2\boldsymbol{1_{T}1{'}_{T}}$, where $\boldsymbol{I_T}$ is a $T\times T$ identity matrix and  $\boldsymbol{1_T}$ is a $T\times 1$ vector of 1's. Using the Sherman-Morrison formula \cite{ShermanMorrisonActual,ShermanMorrison}, we can obtain its inverse as
 $$\boldsymbol{V_i}^{-1} = \frac{1}{\sigma_c^2(\sigma_c^2+T\sigma_{\alpha}^2)}\left[(\sigma_c^2+T\sigma_{\alpha}^2)\boldsymbol{I_T} - \sigma_{\alpha}^2\boldsymbol{1_T1_T}{'}\right].$$
 
This matrix has off-diagonal elements  $\frac{-\sigma_{\alpha}^2}{\sigma_c^2(T\sigma_{\alpha}^2+\sigma_c^2)}$ and diagonal elements  $\frac{(T-1)\sigma_{\alpha}^2+\sigma_c^2}{\sigma_c^2(T\sigma_{\alpha}^2+\sigma_c^2)}$. Due to  the block diagonal structure of $\boldsymbol{V}$, we have
 $$\boldsymbol{Z{'}V}^{-1}\boldsymbol{Z} = \sum_{i=1}^{I}\boldsymbol{Z_i}{'}\boldsymbol{V_i}^{-1}\boldsymbol{Z_i},$$ where $\boldsymbol{Z_i}$ is the $T\times (T+3)$ part of the design matrix corresponding to cluster $i$.
 We can then rewrite \begin{equation}
 \boldsymbol{Z_i}{'}\boldsymbol{V_i}^{-1}\boldsymbol{Z_i} = \frac{1}{\sigma_c^2(\sigma_c^2+T\sigma_{\alpha}^2)}\left[(\sigma_c^2+T\sigma_{\alpha}^2)\boldsymbol{Z_i}{'}\boldsymbol{Z_i} - \sigma_{\alpha}^2\boldsymbol{Z_i}{'}\boldsymbol{1_T1_T}{'}\boldsymbol{Z_i}\right].
 \end{equation} We then use block matrix inversion techniques to solve for the submatrix corresponding to the coefficients corresponding to the treatment and interaction effects. A full derivation is provided in Appendix A. 

\section{Illustrative Examples}
\label{Section:DesignIssues}
Although we derived closed form solutions for the variance and covariances of the treatment and interaction effect estimates, there is no simple, intuitive expression for the standard errors nor for power. We therefore use numerical examples to illustrate how power for detecting treatment and interaction effects is affected by design features for SWDs with two treatments. All calculations were performed in R version 3.6.1 \cite{Rstudio} with code available on Github at https://github.com/phillipsundin/SWFD.

Our main concerns are how power is impacted by the assignment of cluster-periods to conditions and by the factorial aspect of designs. We focus attention on cluster-period assignments that are likely to be used in practice. For example, we only consider designs in which a combined condition follows a control condition or a single treatment condition. We do not consider designs in which a cluster can transition from a combined to a single treatment condition, or from one single treatment to the other single treatment, because such designs are vulnerable to contamination  and therefore would not be commonly used. We also focus on designs with relatively few clusters, which is common in practice \cite{FewClusters}.

For simplicity, we use the repeated cross-sectional design for our examples. Power for repeated cross-sectional versus cohort designs is not a focus of our paper and has been addressed by others \cite{CohortCrossSectionalDesign}. In general, cohort designs have higher power than cross-sectional designs \cite{SampleSize_Longitudinal, RepeatedMeasures}. Often, ICCs are reported as the within-time, within-cluster correlation, $\rho_w$; the across-time, within-cluster correlation $\rho_a$ and individual auto-correlation $\pi$ are not always reported. Given this lack of information, it may be sensible to make the simplifying assumption that $\rho_w = \rho_a$, which corresponds to the repeated cross-sectional design.  

\subsection{Two-Treatment Concurrent SWD vs Two Separate Single Treatment SWDs}
Several studies have conducted two separate one-treatment SWD trials \cite{Dementia, TBScreening}. We explore how power might be increased or the required number of clusters reduced by combining two one-treatment trials into one trial with two treatments. 

\begin{figure}[ht] 
	\centering
	\hspace{-1.75cm}
	\begin{subfigure}{0.275\textwidth}
	\includegraphics[width=1\linewidth]{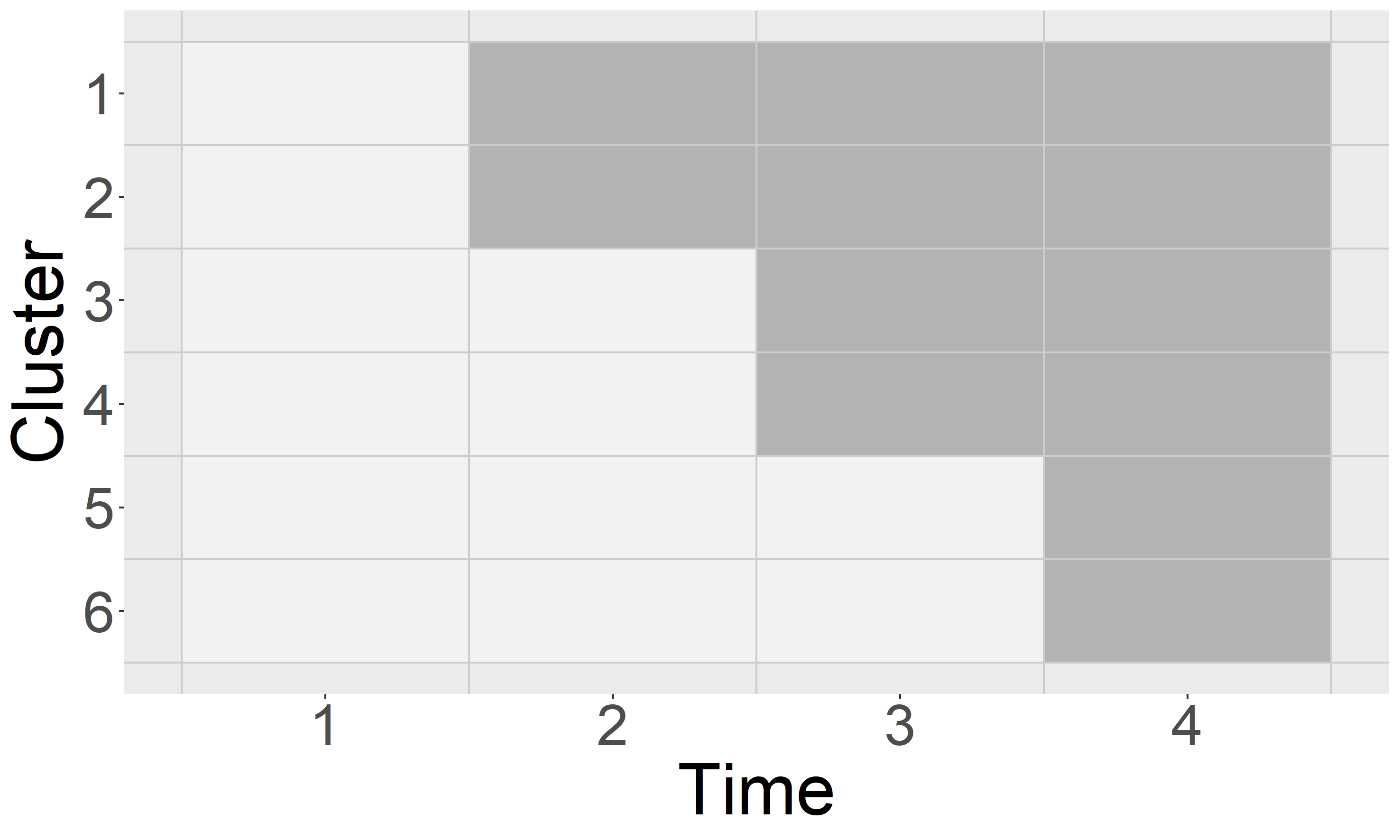}	\includegraphics[width=1\linewidth]{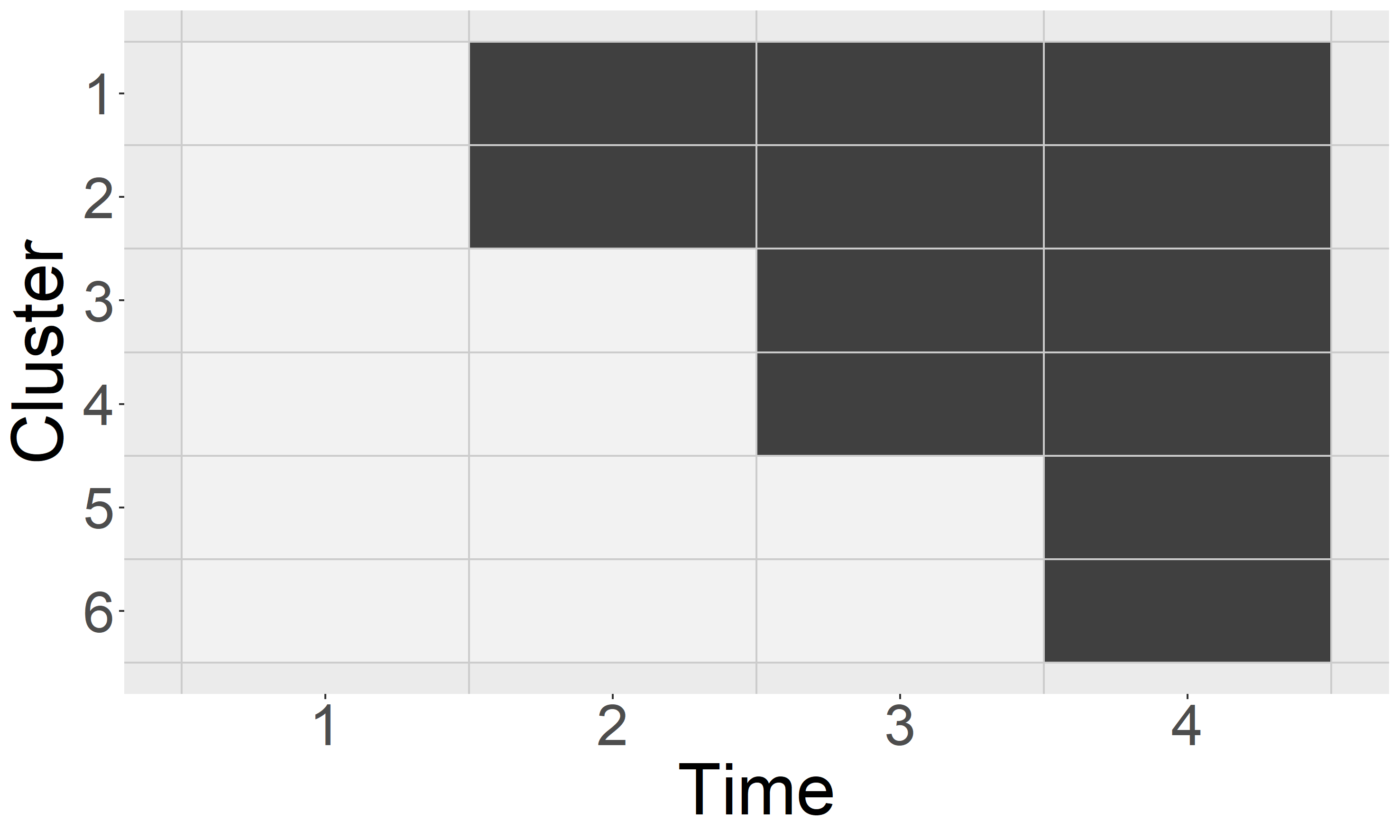}
	\centering
	{{\small (a) Two separate single treatment SWDs}}
\end{subfigure}
\hspace*{0.25cm}
\begin{subfigure}{0.325\textwidth}
\includegraphics[width=1.1\linewidth]{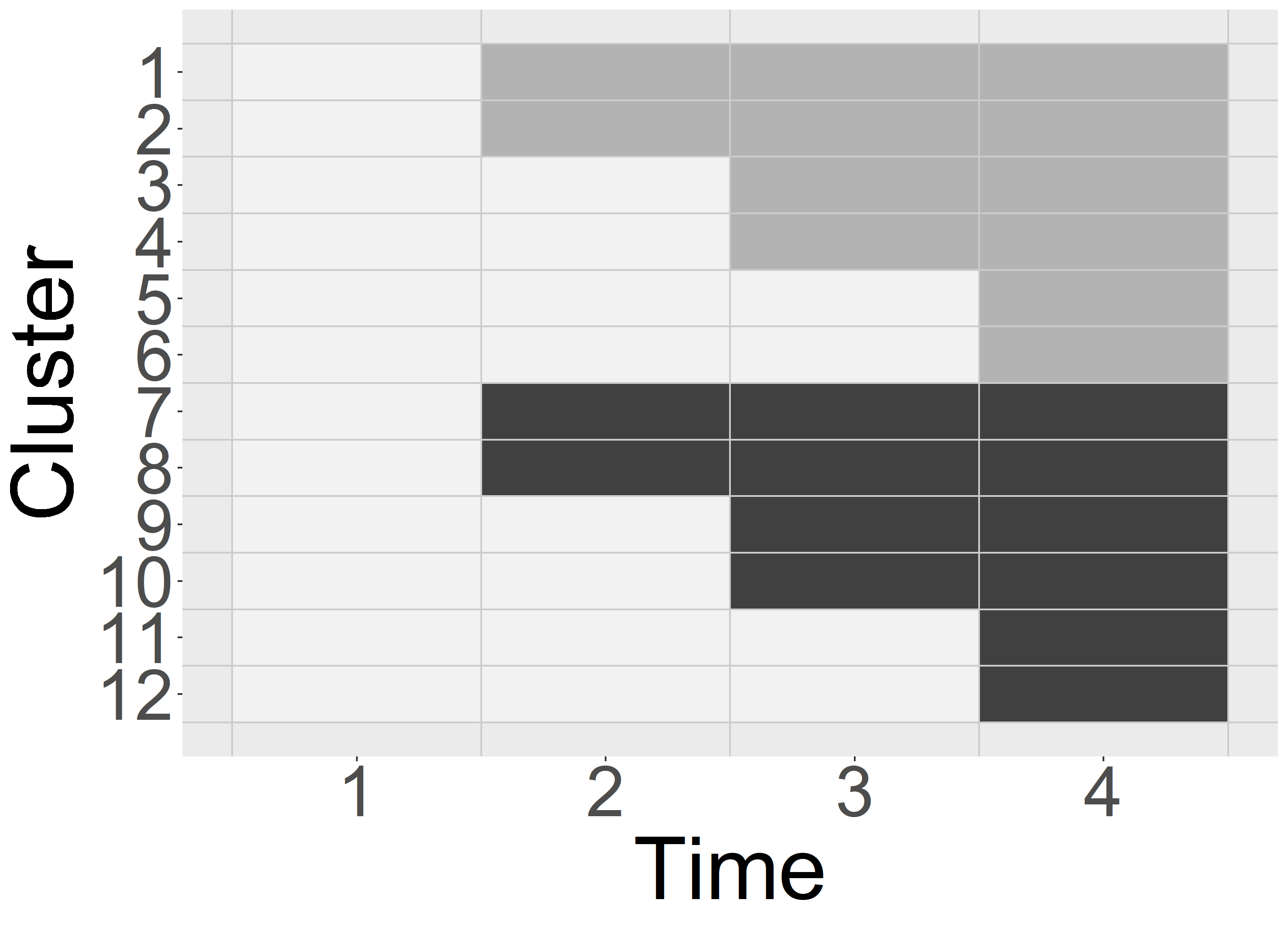}
\centering
{{\small  (b) Concurrent SWD with two treatments, 12 clusters}}  
\end{subfigure}
\hspace*{0.75cm}
\begin{subfigure}{0.325\textwidth}
\includegraphics[width=1.1\linewidth]{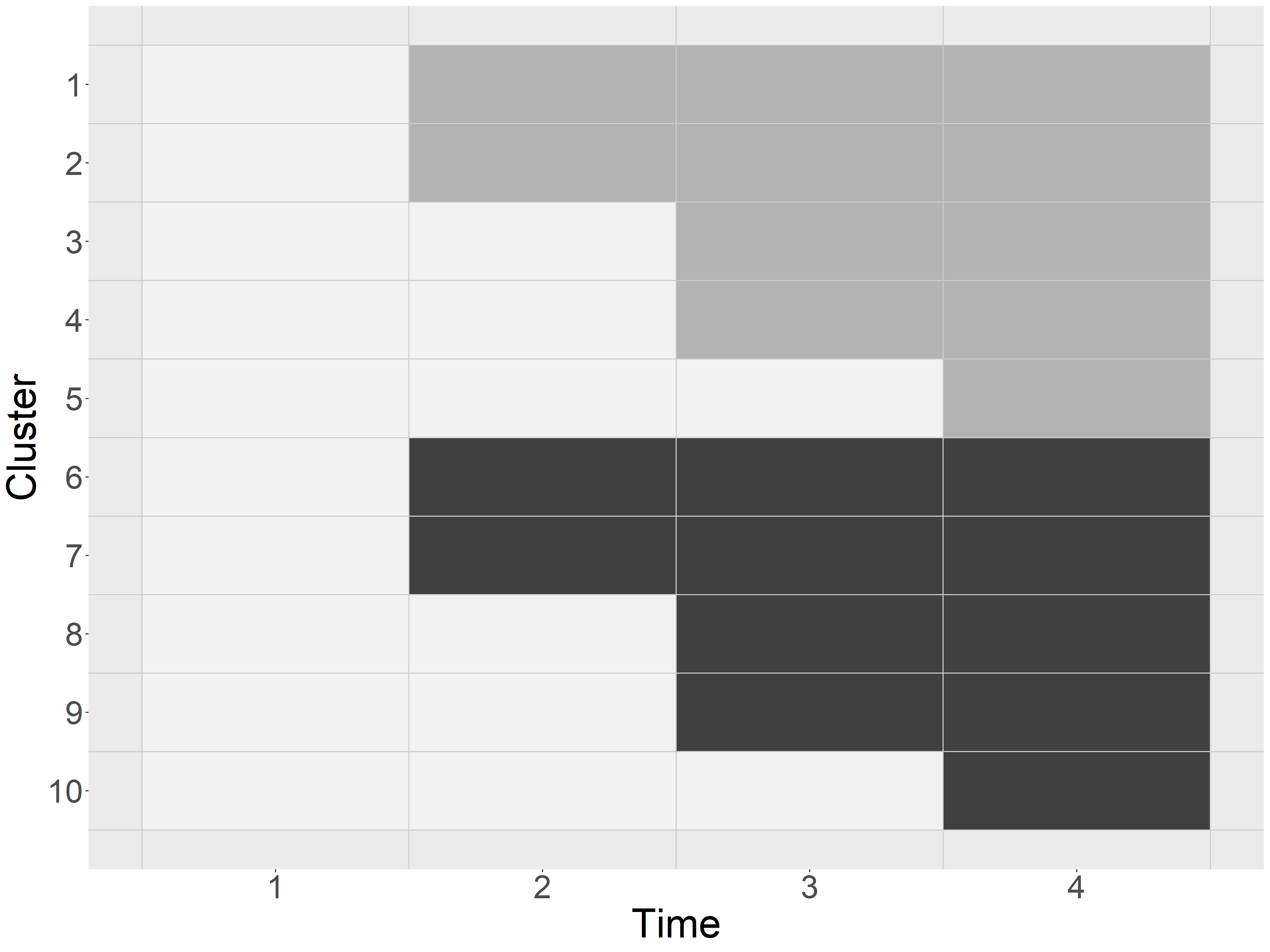}
\centering
{{\small  (c) Concurrent SWD with two treatments, 10 clusters}}  
\end{subfigure}
\caption[SWD]
{\small Examples of two one-treatment SWDs versus concurrent SWDs with two treatments. White cells indicate cluster-periods in the control condition. Light and dark gray cells indicate treatment conditions for treatments 1 and 2, respectively. } 
\label{Figure:2SWD}
\end{figure}

Consider the two one-treatment SWD trials, each with six clusters and four time periods, illustrated in Figure \ref{Figure:2SWD}a. The design in Figure \ref{Figure:2SWD}b combines these into a single 12-cluster trial; such a design has been called a concurrent design \cite{ProposedVariations}. Figure \ref{Figure:2SWD}c shows a similar concurrent design with only 10 clusters. Let $\delta_1$ and $\delta_2$ denote the standardized effect sizes for treatments 1 and 2, respectively. We set $\delta_1 = \delta_2 = 0.4$, representing medium effect sizes \cite{CohensD}. Note that within each design, the power to detect treatment effects 1 and 2 is the same due to symmetry. We specify $N=15$ individuals per cluster-period in a repeated cross-sectional design. Power for detecting a treatment effect in one of the one-treatment SWDs is calculated using model (\ref{Equation:HusseyHughes}); for the concurrent SWDs with two treatments, power is calculated using model (\ref{Equation:NoInteractionModel}). 

\begin{figure}[ht]
	\centering
	\includegraphics[scale=0.3]{"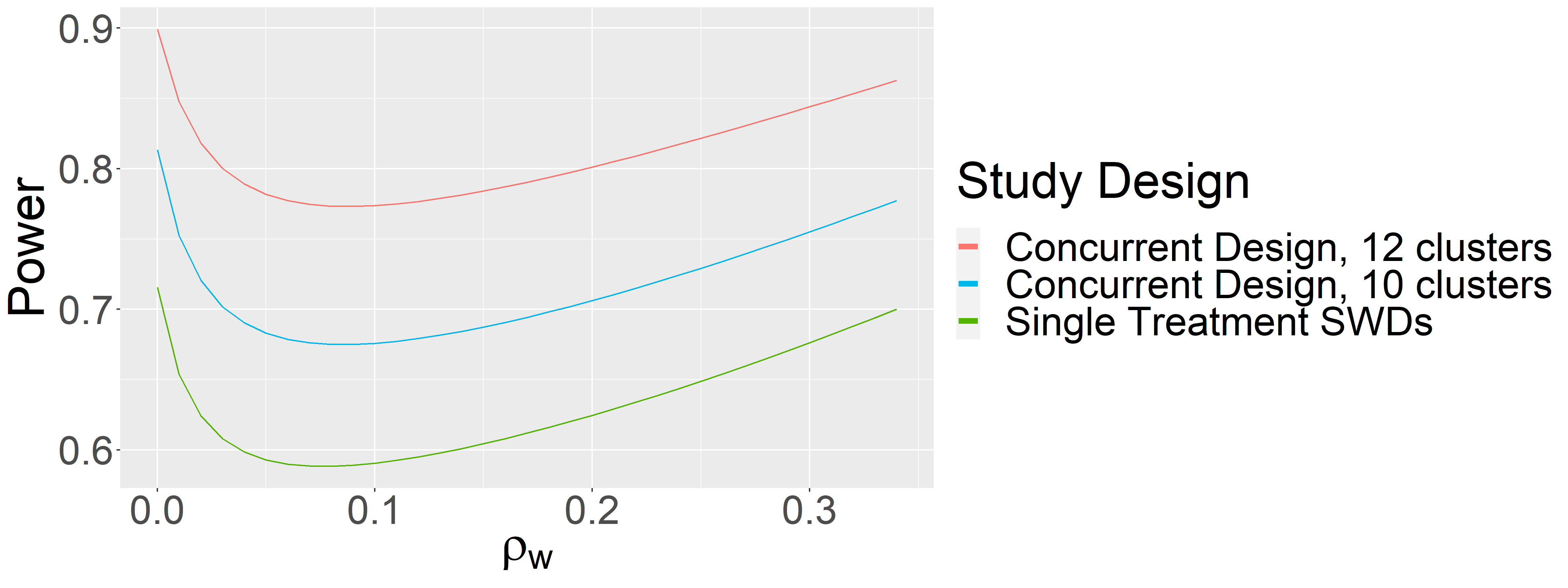"}
	\caption{Comparing power for either treatment effect for two simultaneous one-treatment SWDs, a concurrent SWD with two treatments with 12 clusters, and a concurrent SWD with two treatments with 10 clusters.}
    	\label{Figure:2SWDvsStacked}
\end{figure}

Figure \ref{Figure:2SWDvsStacked} displays power for either treatment effect for the three designs as a function of $\rho_w$. Combining the two one-treatment SWDs into a single concurrent design while maintaining the same number of clusters yields a large gain in power, ranging from 0.14 to 0.20 for the values of $\rho_w$ considered.  The concurrent design with only ten clusters reduces the sample size by about 17$\%$ and has moderately higher power, ranging from 0.08 to 0.11. Further reducing the number of clusters while maintaining symmetry would yield lower power than the single treatment SWDs in this example.

\begin{figure}[ht]
	\centering
	\includegraphics[scale=0.3]{"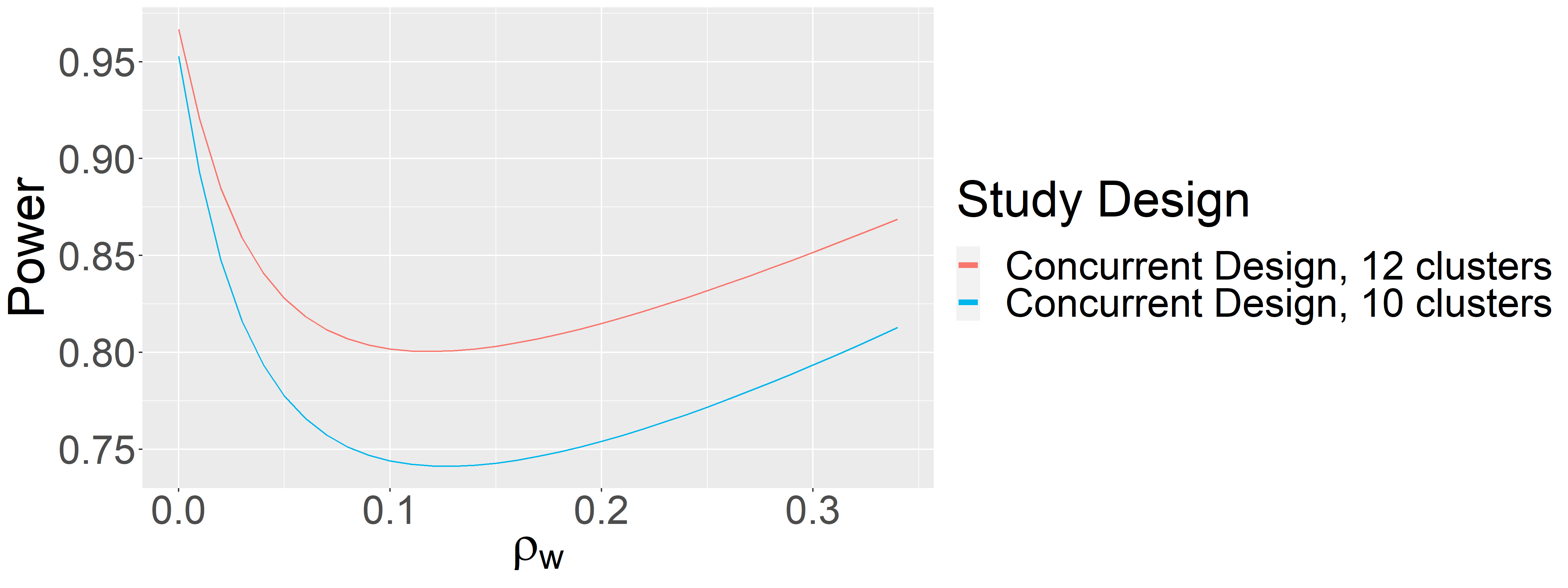"}
	\caption{Comparing power for detecting a difference between two treatments for a 12-cluster concurrent SWD and a 10-cluster concurrent SWD.}
    	\label{Figure:31TreatmentComparison}
\end{figure}

In addition to power gains and reduced sample size requirements, another advantage of including two treatments in one SWD is the capability to directly compare the two interventions. This can be accomplished using tests of linear contrasts of treatment effects. A closed form solution for the standard error of such contrasts, which includes a covariance term, can be computed using our methods. Suppose we wish to detect a standardized difference of 0.4 units between the two treatments. Figure \ref{Figure:31TreatmentComparison} displays power for the linear contrast as a function $\rho_w$. The relationship between $\rho_w$ and power for the linear contrast is similar in shape to the relationship between $\rho_w$ and power for detecting main effects. However, the nadir of the power curve is at a higher value of $\rho_w$ (about 0.12) for the linear contrast in this example. 

\subsection{Stepped Wedge Factorial Designs with Additive Treatment Effects}
\begin{figure}[ht] 
	\begin{subfigure}{0.42\textwidth}
		\includegraphics[width=\linewidth]{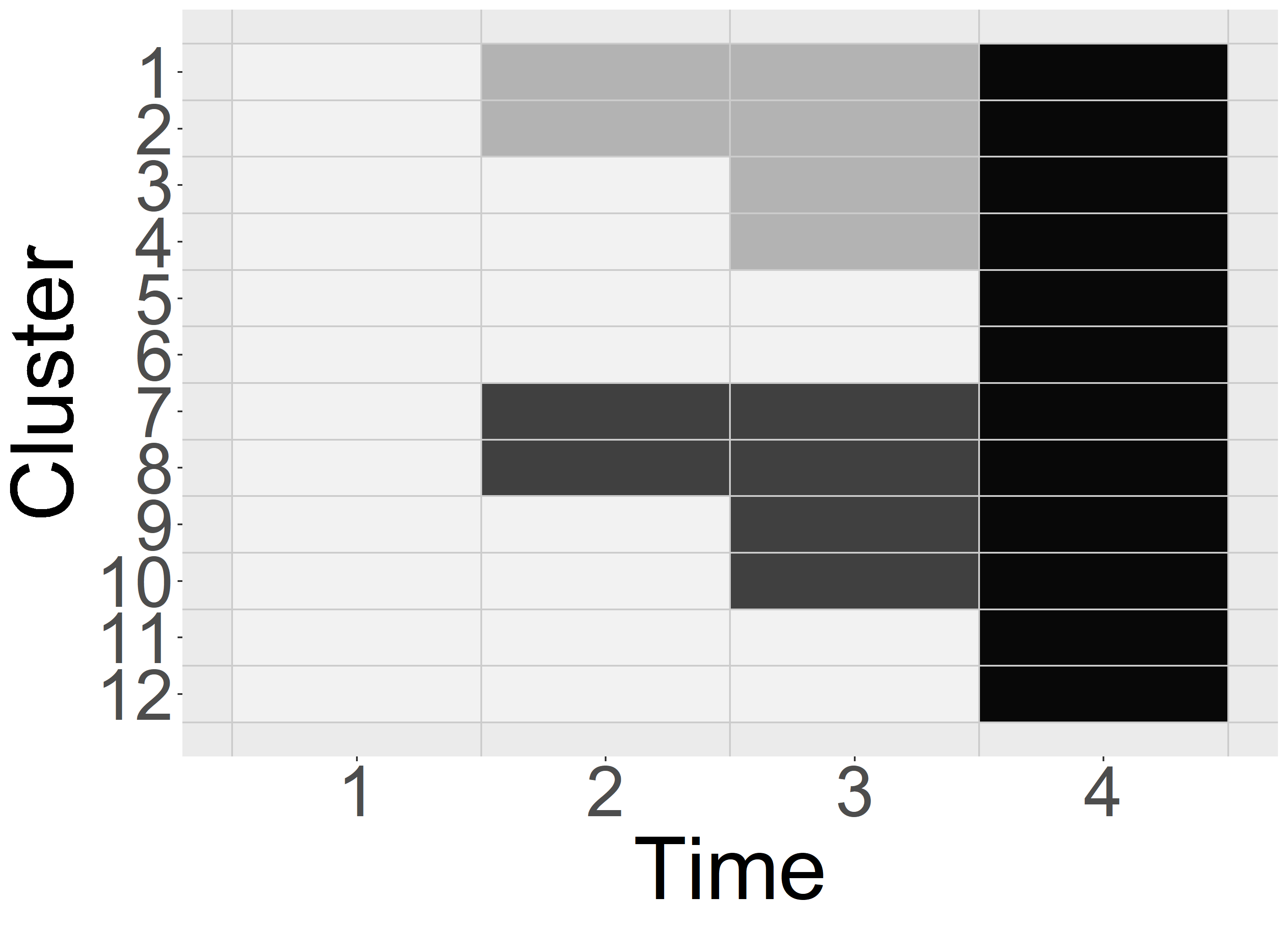}
		\centering
		{{\small (a) "Late" factorial design with 12 clusters in which every cluster receives the combined condition in the last (fourth) period}}   \label{SWFD_StackedCombined}
	\end{subfigure}\hspace{0.75cm}
	\begin{subfigure}{0.5\textwidth}
		\includegraphics[width=1.2\linewidth]{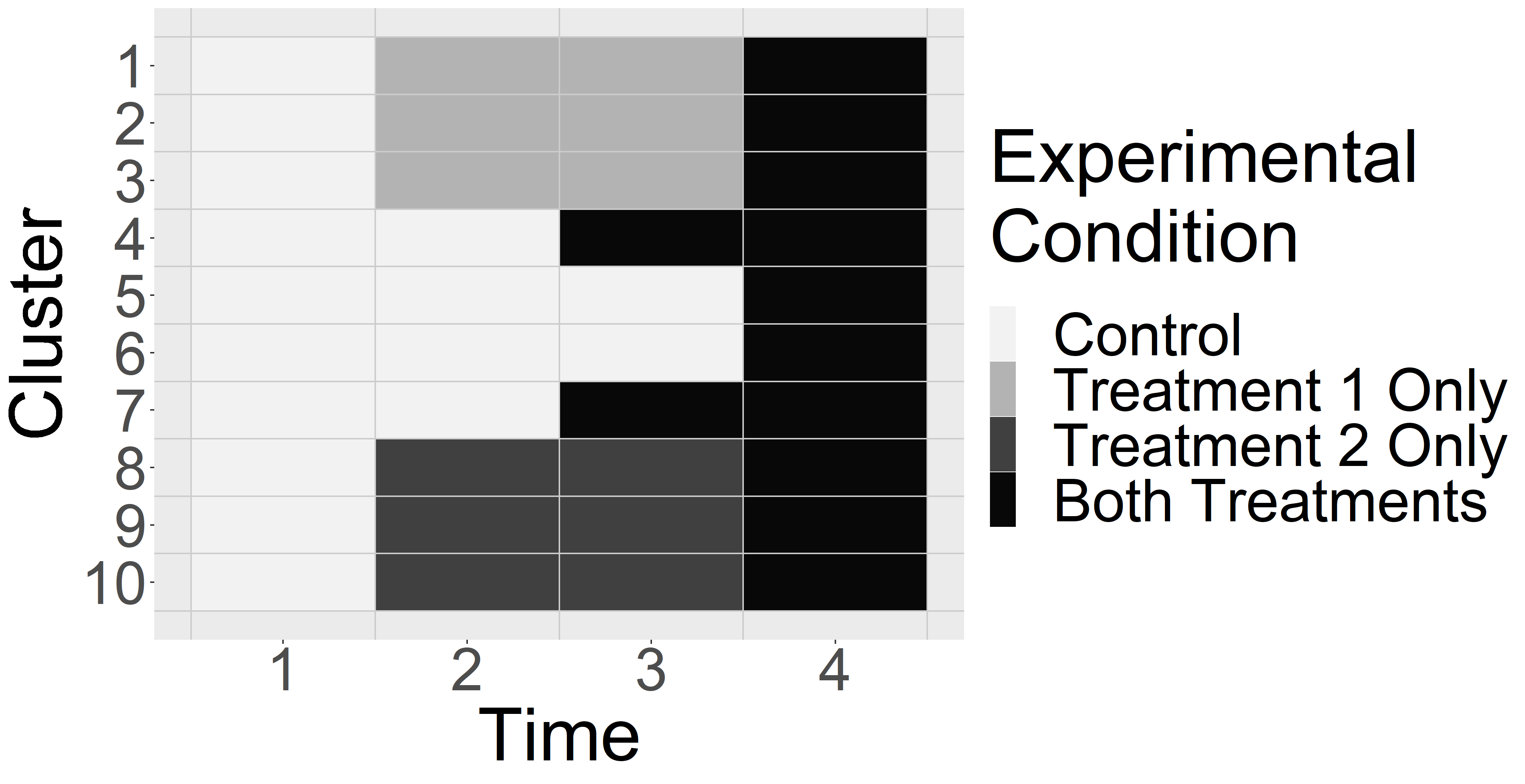}
		\centering
		{{\small (b) "Earlier" factorial design with 10 clusters, in which two clusters begin receiving the combined condition in the third period}}    
		\label{SWFD_10Clusters}
	\end{subfigure}
	\caption[Stepped wedge factorial design examples.]
	{\small Stepped Wedge Factorial Design Examples.} 
	\label{Figure:SWFD}
\end{figure}

We now consider using designs with a combined condition to estimate additive treatment effects. We contrast the 12-cluster concurrent design in Figure \ref{Figure:2SWD}b with SWFDs that feature some cluster-periods receiving both treatments simultaneously. While many designs are possible, for illustrative purposes we consider the two designs in Figure \ref{Figure:SWFD}. Figure \ref{Figure:SWFD}a shows a 12-cluster ``late" factorial design in which all clusters transition to the combined condition for the last period. Figure \ref{Figure:SWFD}b shows an ``earlier" factorial design with only ten clusters that introduces the combined condition earlier. Both designs feature six cluster-periods in each single treatment condition and twelve cluster-periods in the combined condition, representing equal intervention resources. Treatment effects are assumed to be additive. Other design aspects, including repeated cross-sectional observations, moderate effect sizes ($\delta_1 = \delta_2 = 0.4$) and $N=15$ for each cluster-period, are assumed to be the same.

\begin{figure}[ht]
	\centering
	\includegraphics[scale=0.32]{"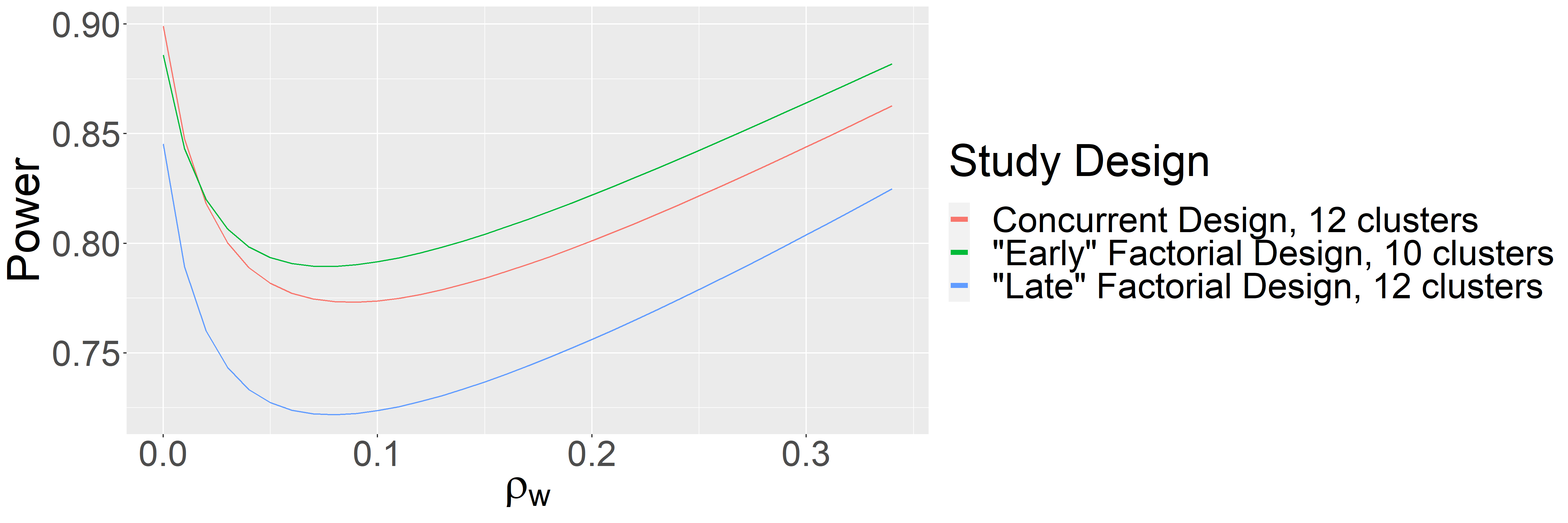"}
	\caption{Comparing power for a 12-cluster concurrent design, a factorial design with 10 clusters, and a factorial design with 12 clusters.}
	\label{Figure:Section32Results}
\end{figure}

Figure \ref{Figure:Section32Results} compares power for the three designs (Figures \ref{Figure:2SWD}b, \ref{Figure:SWFD}a and \ref{Figure:SWFD}b) as a function of $\rho_w$. For $\rho_w < 0.02$, the 12-cluster concurrent design and the 10-cluster ``earlier" factorial design have similar power while the 12-cluster ``late" factorial design has the lowest power. For $\rho_w > 0.02$, the power of the  10-cluster ``earlier" design is highest; the 12-cluster ``late" design continues to have the lowest power. 

This example illustrates several points.  First, when the treatment effects are additive, including a combined condition has the potential to increase efficiency and reduce the number of clusters required to achieve a desired level of power. However, the timing of transitions to the combined condition matters. Simply assigning all clusters to the combined condition for the last period reduced power. Rather, the combined condition needed to be introduced in an earlier period to realize efficiency gains. Furthermore, the value of the ICC is also important. Efficiency gains from using a SWFD were only realized when the ICC exceeded a certain value. A similar phenomenon has been found for a one-treatment SWD compared to a  parallel cluster randomized trial \cite{SampleSizeUnifiedApproach}.

\begin{figure}[ht]
	\centering
	\includegraphics[scale=0.3]{"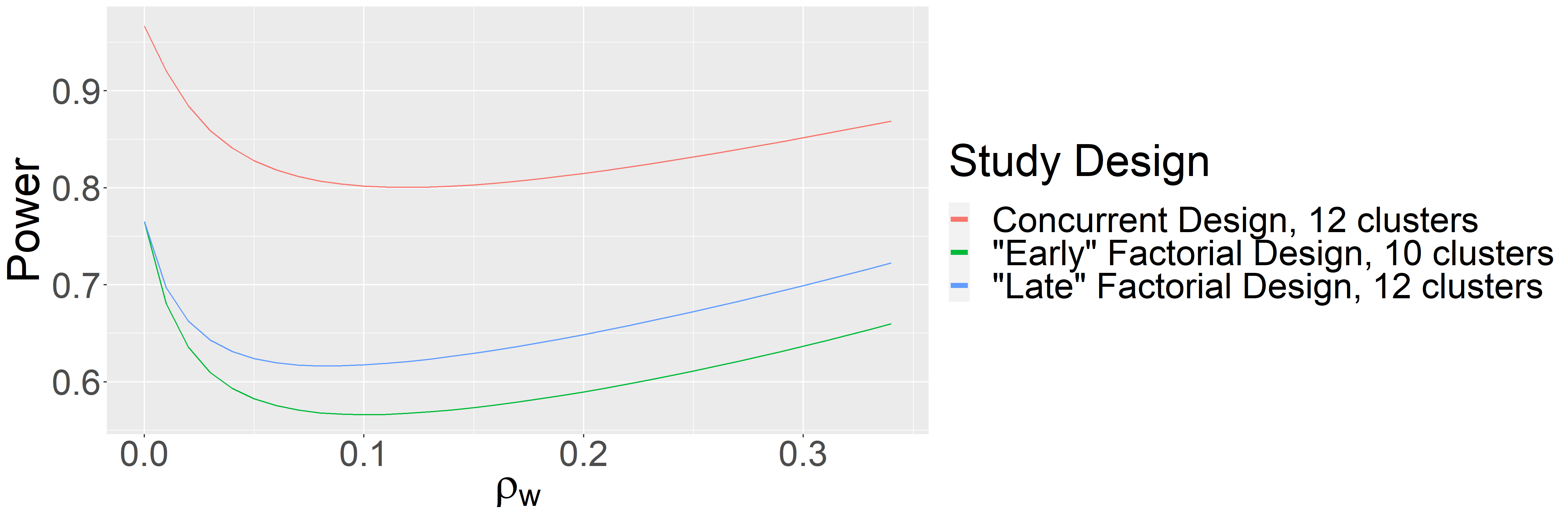"}
	\caption{Comparing power for detecting a difference between two treatments for the concurrent and factorial designs.}
    	\label{Figure:32TreatmentComparison}
\end{figure}

We can also compare the two treatments directly using the SWFD. Suppose we wish to detect a standardized difference of 0.4 between the two treatment conditions. Figure \ref{Figure:32TreatmentComparison} shows that is this example, the concurrent design has the highest power for detecting this contrast across all values of $\rho_w$. For low values of $\rho_w$, the two factorial designs have similar power, but for $\rho_w > 0.10$, the design with 12 clusters has about 7\% higher power compared to the 10-cluster design. 

This example shows that for hypotheses comparing treatments, introducing a combined condition is likely to reduce power compared to a concurrent design. Introducing a combined condition decreases both the variances and covariance  of the two treatment effect estimates; however, the covariance tends to decrease at a much faster rate than the variances as the number of cluster-periods in the combined condition is increased. Because $\text{Var}(\hat{\theta}_1 - \hat{\theta}_2) = \text{Var}(\hat{\theta}_1) + \text{Var}(\hat{\theta}_2) - 2\text{Cov}(\hat{\theta}_1,\hat{\theta}_2)$, the higher covariance values in the concurrent design will result in a smaller overall variance for the contrast, and thus higher power, compared to most SWFDs. This result may seem counterintuitive, since factorial designs are associated with efficiency gains for testing linear contrasts in the context of independent units of randomization. However, note that we still see efficiency gains for testing main effects with this factorial design.

\subsection{Stepped Wedge Factorial Designs with Interaction Effect}
\label{Section:IntroducingInteraction}

\begin{figure*}[ht]
	\centering
	\begin{subfigure}[b]{0.475\textwidth}   
		\centering 
		\includegraphics[width=\textwidth]{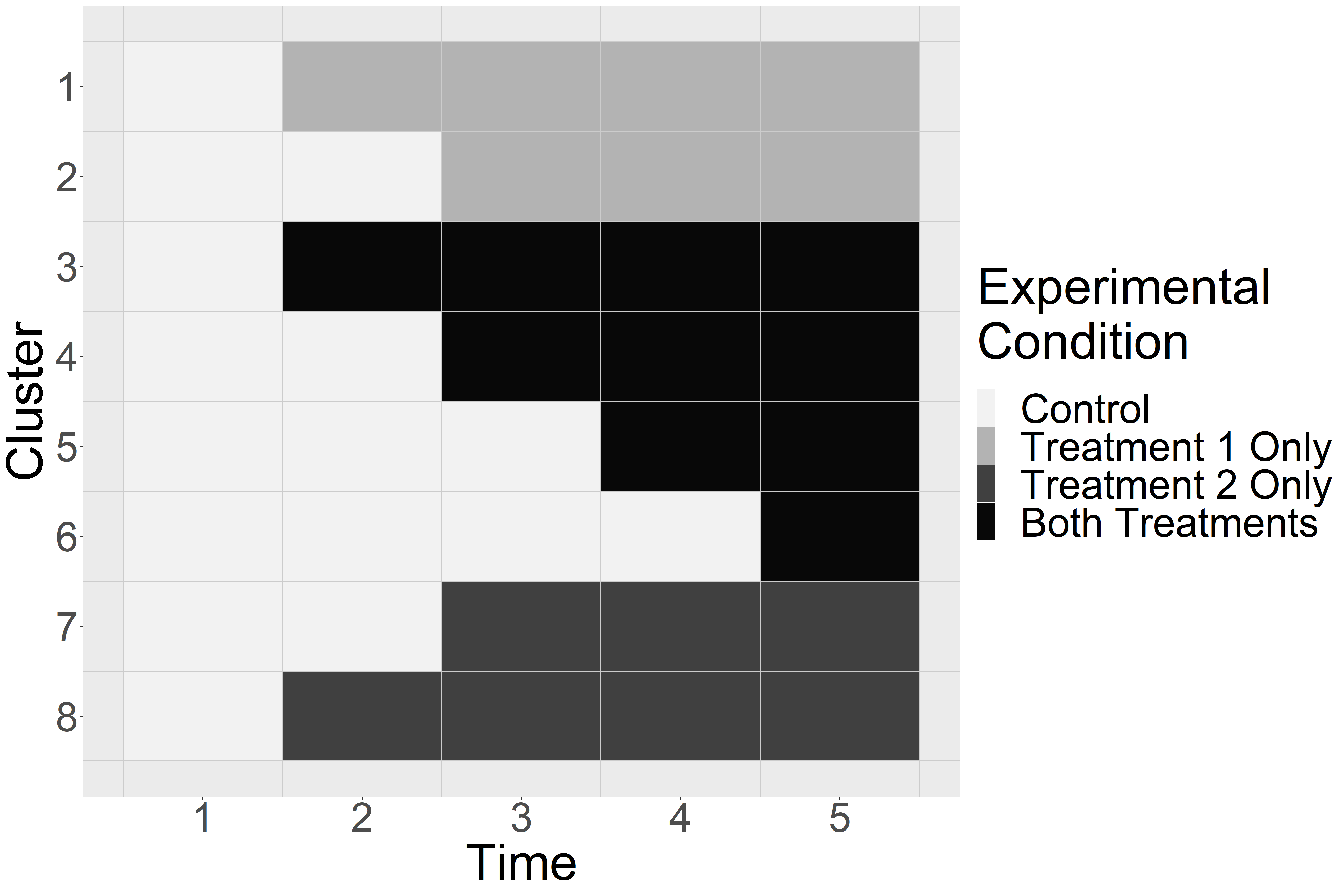}
		{{\small Design \#1}}    
	\end{subfigure}
	\hfill
	\begin{subfigure}[b]{0.475\textwidth}  
		\centering 
		\includegraphics[width=\textwidth]{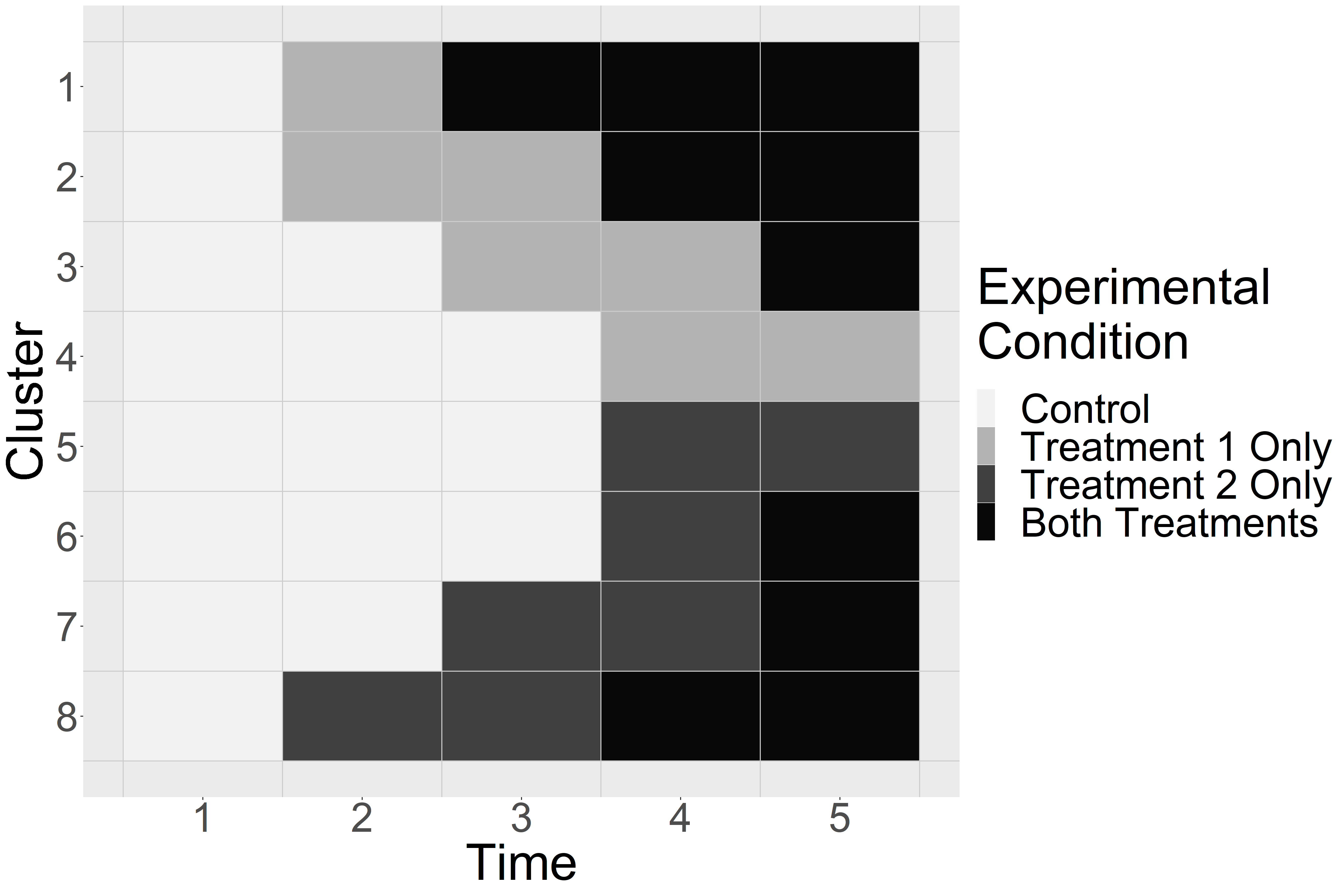}
		{{\small Design \#2}}    
	\end{subfigure}
	\vskip\baselineskip
	\begin{subfigure}[b]{0.475\textwidth}   
		\centering 
		\includegraphics[width=\textwidth]{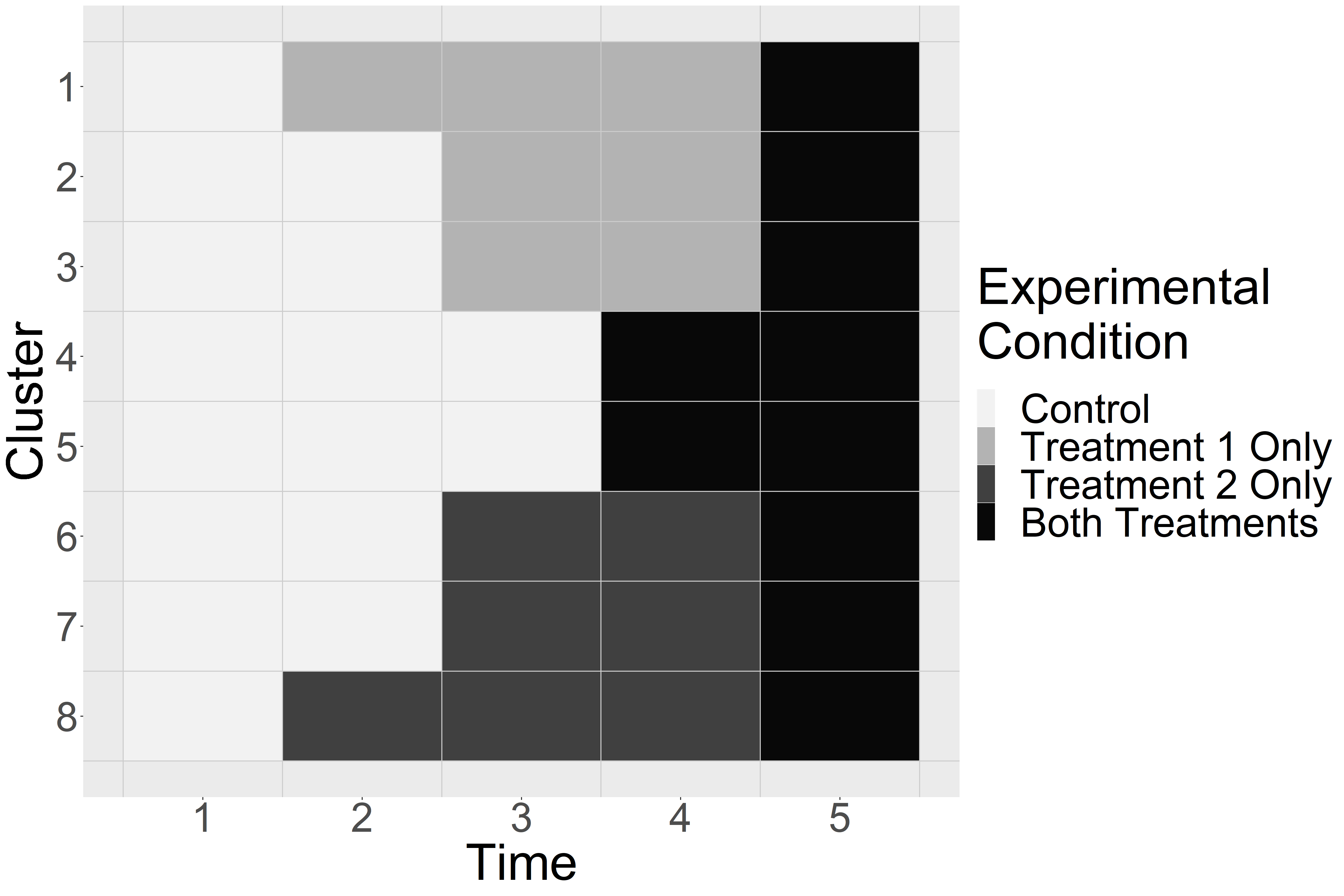}
		{{\small Design \#3}}    
	\end{subfigure}
\hfill
	\begin{subfigure}[b]{0.475\textwidth}
		\centering
		\includegraphics[width=\textwidth]{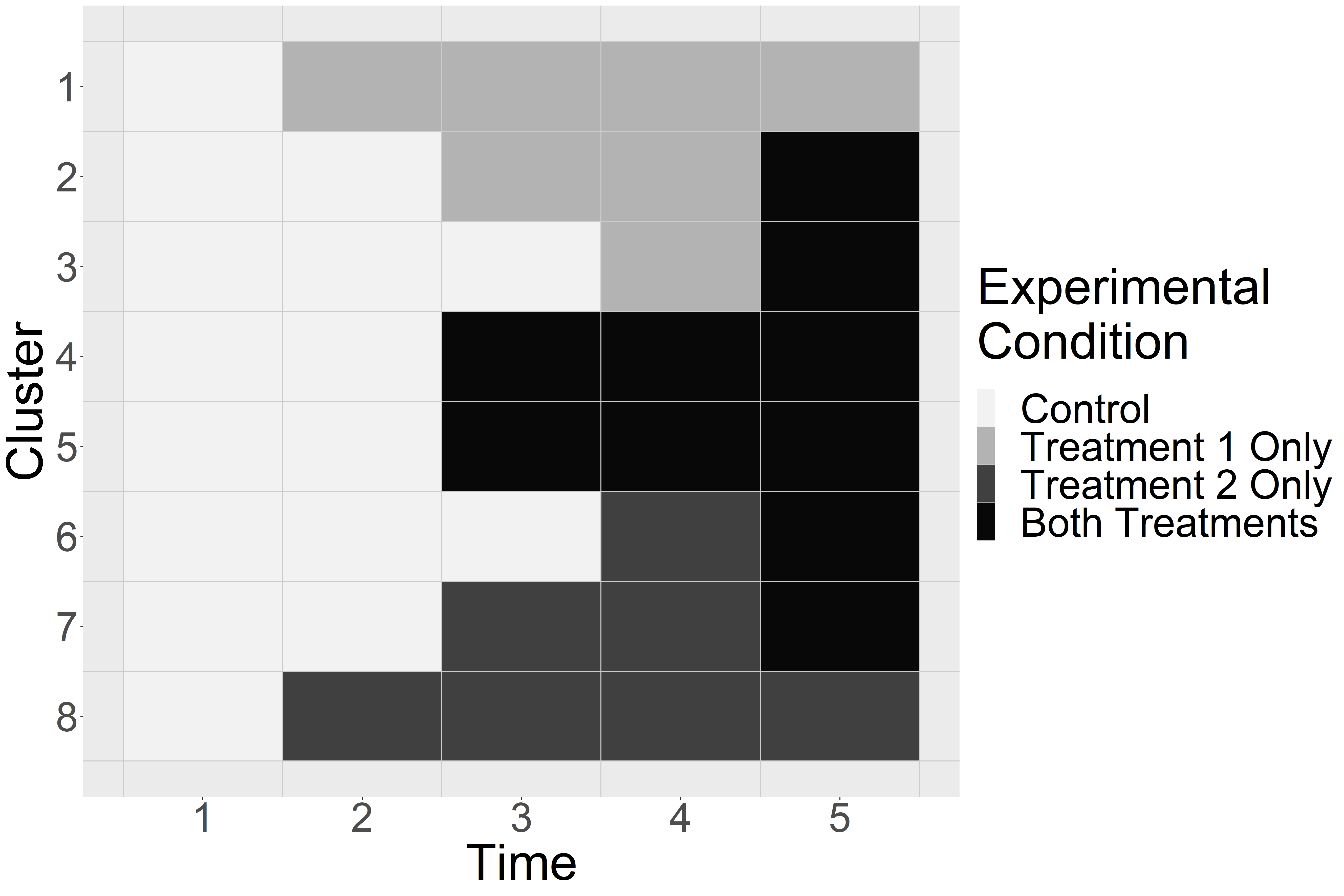}
		{{\small Design \#4}}    
	\end{subfigure}
	\caption[Variations of Stepped Wedge Factorial Design ]
	{\small Varying sequences in a stepped wedge factorial design.} 
	\label{Figure:VariationsSWD_Figure}
\end{figure*}

We now consider designs in which detecting an interaction is of interest. Consider the four designs in Figure \ref{Figure:VariationsSWD_Figure}. All have eight clusters, as well as seven cluster-periods in treatment 1 only, seven cluster-periods in treatment 2 only, and ten cluster-periods in the combined condition. Design \#1 is essentially a concurrent design in which each cluster transitions from control to one condition only during the study, including the combined condition. Design \#2 has three clusters transitioning from control to treatment 1 to combined, three clusters transitioning from control to treatment 2 to combined, and two clusters transitioning from control to a single treatment condition relatively late in the study. Design \#3 has six of the eight clusters transition to a single treatment in classic stepped wedge style, and all clusters transition to the combined condition.  Design \#4 is similar but includes two clusters that never transition to the combined condition as well as combined condition clusters that start earlier. 

We again assume a repeated cross-sectional design with 15 individuals per cluster-period. We calculate power to detect a standardized effect size of 0.60 for each main and interaction effect. Because this represents a relatively large interaction effect, this example illustrates how design choices impact power for an interaction effect. 

\begin{figure}[ht]\hspace{-0.5cm} 
	\begin{subfigure}{0.5\textwidth}
		\caption{Main Effects}
	\label{Figure:PowerSWFDComparison}
	\centering
	\includegraphics[scale=0.23]{"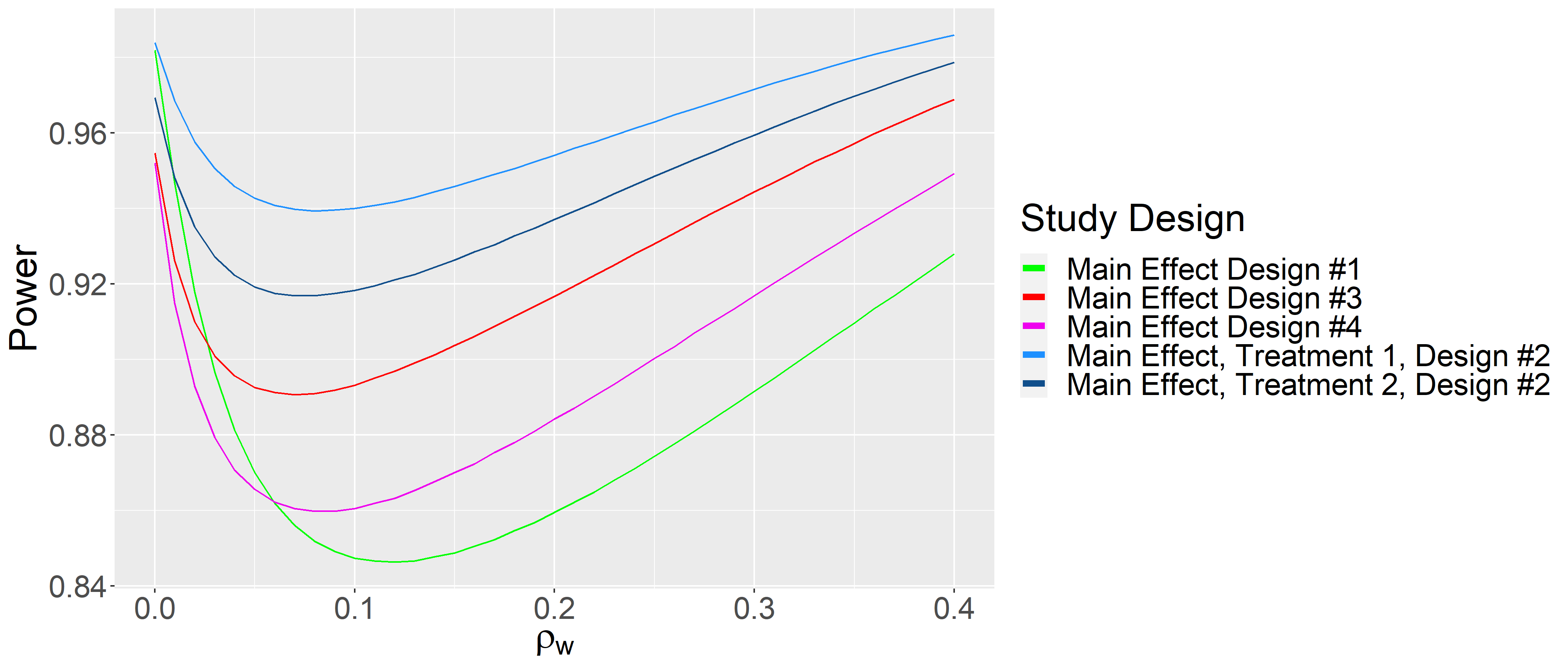"}
	\end{subfigure}\hspace{.95cm}
	\begin{subfigure}{0.5\textwidth}
	\caption{Interaction Effect}
	\centering
	\includegraphics[scale=0.23]{"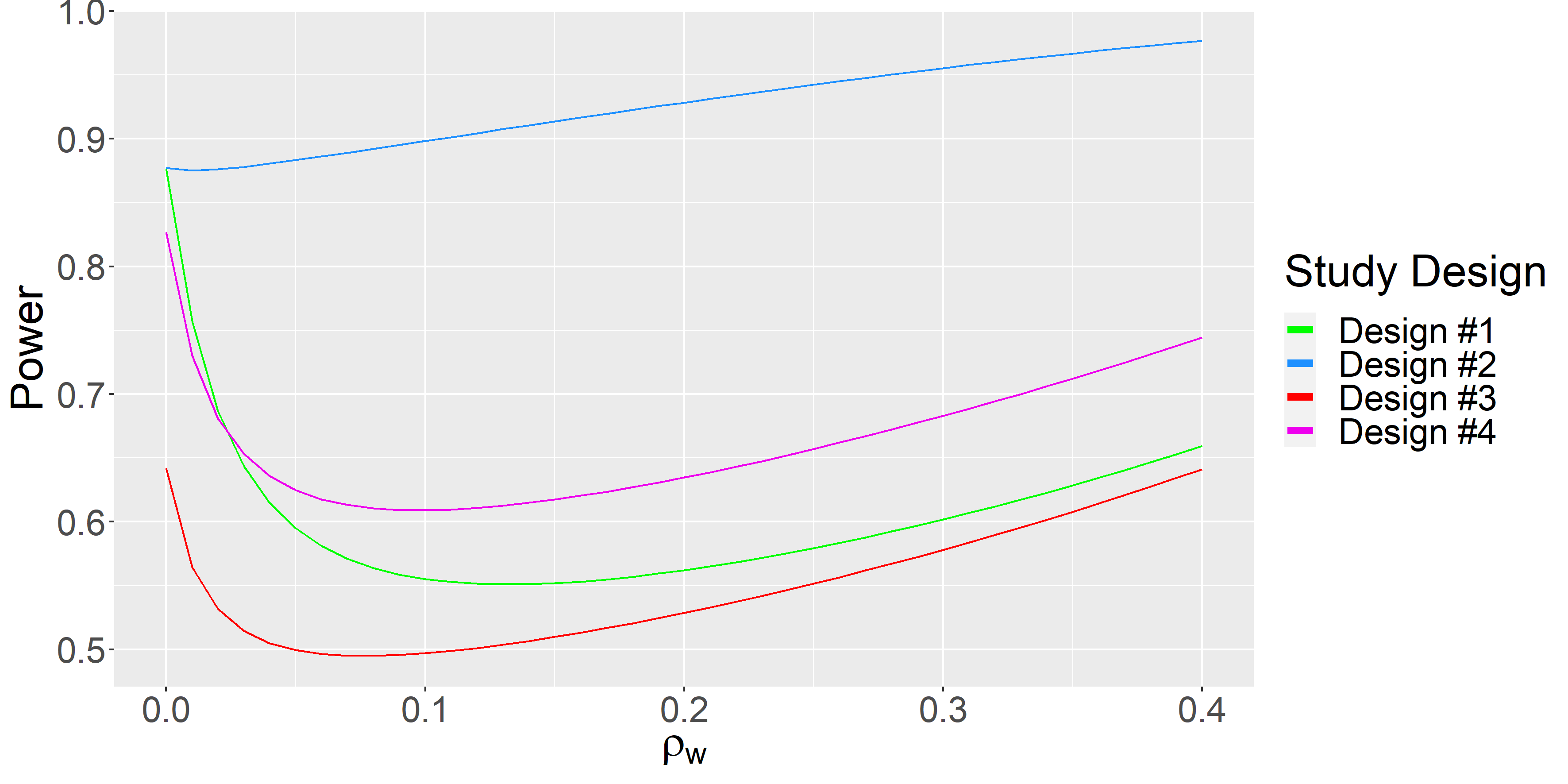"}
	\end{subfigure}
	\caption[Comparing Main and Interaction Effects for Four SWFDs]
	{\small Comparing main and interaction effects for SWFDs with different sequences.} 
	\label{Figure:InteractionEffect}
\end{figure}

Figure \ref{Figure:InteractionEffect} presents power as a function of $\rho_w$ for the four designs. Note that Designs \#1, \#3, and \#4 are symmetric in treatments 1 and 2 and thus power for these two treatment effects is the same in these designs. Design \#2 is close to symmetric, but symmetry may be hard to achieve with a small number of clusters. We focus first on power for main effects, displayed in Figure \ref{Figure:InteractionEffect}a. For all values of $\rho_w$, Design \#2 has the highest power for detecting main effects. In this design, power for treatment 1 is slightly higher than that for treatment 2. This result is attributable to a more balanced sequencing of treatment 1 over time compared to treatment 2, as the majority of cluster-periods with only treatment 2 are administered later in the study. For values of $\rho_w<0.06$, Design \#4 has the lowest power; for $\rho_w>0.06$, Design \#1 (concurrent) has the lowest power. Focusing on power for the interaction effect, displayed in Figure \ref{Figure:InteractionEffect}b, Design \#2 has by far the highest power, while Design \#3 has the lowest power. Design \#1 also has relatively low power.

Design \#2 is clearly superior for detecting both main and interaction effects. In Design \#2, clusters transition between conditions more than any other design. When there are more transitions, within-cluster comparisons are increased, and thus power to detect effects is increased. Design \#1 does not have any clusters transition to different treatment conditions after the first transition from the control condition, and thus suffers from the lowest power for main effects. For the interaction effect, Design \#3 suffers from the lowest power.  This can be partially attributed to confounding with time, as all of the clusters are in the combined condition during the last time period, and there are only two cluster-periods in the combined condition before the last time period. 

This example illustrates that to power on the interaction term, designs should include (1) clusters that transition from control to single treatment to the combined condition and (2) transitions to the combined condition should occur relatively early in the study. Our examples also show that power for both main and interactions effect in SWFDs has a non-monotone relationship with $\rho_w$, with the lowest power for values of $\rho_w$ between 0 and 0.2, depending on the particular design. 

We note that in all of our examples, the power curves have had a non-monotone relationship with $\rho_w$, such that power is a decreasing function of $\rho_w$ at lower values of $\rho_w$ and an increasing function at higher values. This is consistent with power curves for stepped wedge designs with only one treatment \cite{ReduceSampleSize, SampleSizeUnifiedApproach, SampleSizeCalculationSWD}.

\section{Discussion}
\label{Section:Discussion}
In this paper, we present power calculation methods for stepped wedge trials that involve two treatments, with and without a combined condition or interaction. Designs with two treatments are increasingly being used in practice, despite a paucity of literature on their statistical design and analysis. 

We focus on studies that include a relatively small number of clusters, which is common for stepped wedge trials \cite{FewClusters}. In our examples, it was not possible to explore all possible design options. However, the examples demonstrate several principles.  We show that a concurrent design, in which two one-treatment stepped wedge trials are conducted as a single study, is more efficient than two separate one-treatment studies, and we enable power calculations for such studies. In concurrent designs, cluster-periods in the control condition perform ``double duty"  by serving as controls for both treatment conditions. Such trials are essentially three-arm trials in which two interventions are each compared to a control condition. 

When designing trials with two treatment conditions, investigators may wish to consider the need to control the experimentwise type I error rate. We note that a concurrent stepped wedge design is similar to a one-way analysis of variance design in which multiple treatment groups are each compared to a common control group. In this setting, Dunnett's method may be used for experimentwise type I error rate control \cite{Dunnett}.

Our examples illustrate that factorial stepped wedge designs that include cluster-periods in a combined condition can increase power substantially compared to concurrent designs when treatment effects are additive. However, power may end up being inadequate if an interaction between treatments is manifested in the study but was not taken into account in power calculations. One approach for guarding against this eventuality is to conduct sensitivity analyses that assume some interaction between treatments when designing the study. Our proposed power calculation methods can be used for this purpose.

Our models allow for an interaction between two treatments in a stepped wedge design. In many studies, detecting an interaction effect may be of scientific interest. The presence of an interaction generally decreases power for detecting main effects in factorial designs \cite{FactorialDiscussion}. In a SWFD where the aims include detecting an interaction effect, treatment sequencing is critical. Our examples show that in general, designs in which clusters transition from control to single treatment to combined treatment will be more powerful than designs in which clusters make only one transition, from control to single treatment or control to combined condition. Such multiple-transition designs allow for more within-cluster comparisons, which are a driving factor in power for stepped wedge trials in general. 

In this paper, we focus on SWDs with two treatments. However, if every condition can be represented by a binary indicator, the results are generalizable to any number of treatments and two-way interactions. Consider a model with $M$ main effects and $B$ two-way interaction terms. The variance-covariance matrix of the regression coefficients would be a $(M+B)\times(M+B)$ matrix. The elements of this matrix would have the same form as the elements of the $3\times3$ matrix in Appendix A for diagonal and off-diagonal elements for both main and interaction effects. Solving for $[(M+B)\times(M+B)]^{-1}$  would yield the variance-covariance matrix for the estimated coefficients. Note that this approach holds for two-way interactions only; higher-order interactions are not considered. 

We consider continuous outcomes only. Further development is needed for non-continuous outcomes, including binary, survival, categorical and count outcomes. For the nested exchangeable model in Section \ref{Subsection:TimeTrend}, the cluster autocorrelation is constrained to be the same for cluster means across time periods, regardless of the length of time between observing cluster level outcomes. This may not be an accurate assumption, as cluster means observed closer in time may be more correlated than those that are farther apart \cite{AnalysisCRTRepeatedCS}. There are models for one treatment SWDs that allow the correlation between cluster means to decay over time \cite{GEE_DecayCorrelation, DecayCorrelation,MixedEffectsSWDOverview}. For linear mixed models with a decaying correlation structure, the covariance matrix is a Toeplitz matrix and requires the use of the Trench algorithm to numerically invert \cite{DecayCorrelation}. We did not include this feature in our work here as we focused on the derivation of closed form variances and covariances of treatment and interaction effects. In this paper, we only consider complete SWDs. Incomplete designs, in which for some periods, data are not collected from some clusters, have been addressed for stepped wedge trials with a single treatment \cite{GenericFramework, InformationContent_ImplementationPeriods}. Finally, we have assumed that treatment effects are instantaneous and do not consider any delay in treatment effects, which have been considered for SWDs with a single treatment \cite{HusseyHughes, MixedEffectsSWDOverview, CurrentIssues}. Future work could explore how delays in one or both treatment effects may impact power of main and interaction effects.

\section*{Appendix A: Derivation of Standard Errors of Treatment Effect Estimates}

\setlength{\abovedisplayskip}{2.5pt}
\setlength{\belowdisplayskip}{2.5pt}
\setlength{\abovedisplayshortskip}{2.5pt}
\setlength{\belowdisplayshortskip}{2.5pt}


The variance-covariance matrix of the fixed effects in a linear mixed model is found by taking the inverse of $\boldsymbol{Z^{'}V}^{-1}\boldsymbol{Z}$ in which $\boldsymbol{Z}$ is the fixed effects design matrix and $\boldsymbol{V}$ is the variance-covariance matrix of the outcome. In the SWFD model presented in (\ref{Equation:MainSWDModel}), $\boldsymbol{Z}$ is the $IT \times (T+3)$ design matrix and $\boldsymbol{V}$ is the $IT \times IT$ variance-covariance matrix of the outcome. We limit ourselves to finding closed form expressions only for the variance-covariance matrix of the treatment effect estimates, $\hat\theta_1, \hat\theta_2, \text{ and } \hat\theta_3$. We first find an expression for $\boldsymbol{Z^{'}V}^{-1}\boldsymbol{Z}$ and then use block matrix inversion techniques to get the desired elements of $(\boldsymbol{Z^{'}V}^{-1}\boldsymbol{Z})^{-1}$. 

\subsection*{Defining the Precision Matrix}
We focus on the repeated cross-sectional model. For the nested exchangeable and cohort models, the appropriate values for elements of $\boldsymbol{V}$ can be substituted. 

Assuming that clusters are independent, for the repeated cross-sectional model, the matrix $\boldsymbol{V}$ has block diagonal structure with elements $\boldsymbol{V_i} = \sigma_c^2\boldsymbol{I_T} + \sigma_{\alpha}^2\boldsymbol{1_{T}1{'}_{T}}$, where $\boldsymbol{I_T}$ is a $T\times T$ identity matrix and  $\boldsymbol{1_T}$ is a $T\times 1$ vector of 1's. Using the Sherman-Morrison formula \cite{ShermanMorrisonActual,ShermanMorrison} for the inverse of a matrix of this form, we obtain  $$\boldsymbol{V_i}^{-1} = \frac{1}{\sigma_c^2(\sigma_c^2+T\sigma_{\alpha}^2)}\left[(\sigma_c^2+T\sigma_{\alpha}^2)\boldsymbol{I_T} - \sigma_{\alpha}^2\boldsymbol{1_T1^{'}_T}\right].$$ 
Due to the the block diagonal structure of $\boldsymbol{V}$, we can write the precision matrix as
$$\boldsymbol{Z^{'}V}^{-1}\boldsymbol{Z} = \sum_{i=1}^{I}\boldsymbol{Z_i^{'}}\boldsymbol{V_i}^{-1}\boldsymbol{Z_i}.$$ The submatrix $\boldsymbol{Z_i}$ is the $T\times (T+3)$ subset of $\boldsymbol{Z}$ corresponding to cluster $i$. We then can write \begin{equation}
\label{Equation:Submatrix}
\boldsymbol{Z_i^{'}}\boldsymbol{V_i}^{-1}\boldsymbol{Z_i} = \frac{1}{\sigma_c^2(\sigma_c^2+T\sigma_{\alpha}^2)}\left[(\sigma_c^2+T\sigma_{\alpha}^2)\boldsymbol{Z_i^{'}}\boldsymbol{Z_i} - \sigma_{\alpha}^2\boldsymbol{Z_i^{'}}\boldsymbol{1_T1_T^{'}}\boldsymbol{Z_i}\right].
\end{equation}

In the following, the vectors $\boldsymbol{X}, \boldsymbol{W}, \text{ and } \boldsymbol{XW}$ correspond to the columns of the design matrix corresponding to treatment 1, treatment 2, and the interaction term, respectively. Further, let  $\boldsymbol{X_i}$ be the $(T\times1)$ vector that corresponds to cluster $i$, and $\boldsymbol{X_{i,-T}}$ be the $(T-1)\times1$ vector for cluster $i$ that does not include the value of $\boldsymbol{X_i}$ at time $T$. We use similar notation for $\boldsymbol{W_{i}} \text{ and } \boldsymbol{W_{i,-T}}$ for treatment 2, and we use $\boldsymbol{XW_{i}} \text{ and } \boldsymbol{XW_{i,-T}}$ for the interaction term. Using the summation of submatrices in (\ref{Equation:Submatrix}), we can write the precision matrix  $\boldsymbol{Z^{'}V}^{-1}\boldsymbol{Z}$ as a $(T+3)\times(T+3)$ symmetric matrix whose lower triangular elements are
\[\boldsymbol{Z}^{'}\boldsymbol{V^{-1}Z} = \begin{bmatrix}
Tf &  & &  & \\
f\boldsymbol{1_{T-1}} & (f+gT)\boldsymbol{I_{T-1}}  - g\boldsymbol{1_{T-1}1_{T-1}^{'}} &  &   &
\\
y_1  & \sum_{i=1}^{I}\frac{\boldsymbol{X^{'}_{i,-T}}}{\sigma_c^2}  - \sigma_{\alpha}^2h_1\boldsymbol{1_{T-1}^{'}} & l_1 - z_1 & &  \\
y_2 & \sum_{i=1}^{I}\frac{\boldsymbol{W^{'}_{i,-T}}}{\sigma_c^2}  - \sigma_{\alpha}^2h_2\boldsymbol{1^{'}_{T-1}} & q_1 & l_2 - z_2 & \\
y_3 & \sum_{i=1}^{I}\frac{\boldsymbol{(XW)^{'}_{i,-T}}}{\sigma_c^2}  - \sigma_{\alpha}^2h_3\boldsymbol{1^{'}_{T-1}} & q_2 & q_3 & l_3 - z_3
\end{bmatrix}.
\]
\setlength{\abovedisplayskip}{2.5pt}
\setlength{\belowdisplayskip}{2.5pt}
\setlength{\abovedisplayshortskip}{2.5pt}
\setlength{\belowdisplayshortskip}{2.5pt}
The term $\boldsymbol{1_{T-1}}$ is a $(T-1)\times1$ vector of 1's and $\boldsymbol{I_{T-1}}$ is a $(T-1)$ identity matrix, and we define
\begin{equation*} 
\hspace{-0.25cm}a = \frac {1}{\sigma_c^2+T\sigma_{\alpha}^2}, \;\;\; b = \frac{1}{\sigma_c^2}, \;\;\;\; c = ab, \;\;\;\; X^{T} = \sum_{j=1}^T X_{ij}, \;\;\;\; W^{T} =  \sum_{j=1}^T W_{ij}, \;\;\;\; (XW)^{T} =  \sum_{j=1}^T X_{ij}W_{ij},
\end{equation*}
\begin{equation*} 
X^{IT} = \sum_{i=1}^I \sum_{j=1}^T X_{ij}, \;\;\;\; W^{IT} = \sum_{i=1}^I \sum_{j=1}^T W_{ij}, \;\;\;\; (XW)^{IT} = \sum_{i=1}^I \sum_{j=1}^T X_{ij}W_{ij}, \;\;\;\; f = Ia,  \;\;\;\;  g = Ic\sigma_{\alpha}^2,
\end{equation*}
\begin{equation*} \hspace{-0.25cm}  y_1 = aX^{IT},\;\;\;\; y_2 = aW^{IT},\;\;\;\; y_3 = a(XW)^{IT}, \;\;\;\;  h_1 = cX^{IT}\;\;\;\; h_2 = cW^{IT}, \;\;\;\; h_3 = c(XW)^{IT}, \;\;\;\; z_1 = c\sigma_{\alpha}^2\sum_{i=1}^{I}(X^T)^2, \end{equation*}   \begin{equation*}z_2 = c\sigma_{\alpha}^2\sum_{i=1}^{I}\left( W^T \right)^2, \;\;\;z_3 = c\sigma_{\alpha}^2\sum_{i=1}^{I}\left( (XW)^T \right)^2, \;\;\;\; l_1 =bX^{IT}, \;\;\;\;\; l_2 = bW^{IT}, \;\;\;\;\;l_3 = b(XW)^{IT},  \end{equation*} \begin{equation*} q_1 = l_3 - c\sigma_{\alpha}^2\sum_{i=1}^{I}\left(X^T\right)\left(W^T\right), \;\;\;\; q_2  = l_3 - c\sigma_{\alpha}^2\sum_{i=1}^{I}\left((XW)^T\right)\left(X^T\right), \;\;\ q_3 =l_3 - c\sigma_{\alpha}^2\sum_{i=1}^{I}\left((XW)^T\right)\left(W^T\right).  \end{equation*}   The terms $q_2$ and $q_3$ make use of the relationship $X_{ij}^2 = X_{ij} \text{ and } W_{ij}^2 = W_{ij}$. \setlength{\abovedisplayskip}{0pt}
\setlength{\belowdisplayskip}{0pt}
\setlength{\abovedisplayshortskip}{0pt}
\setlength{\belowdisplayshortskip}{0pt} \subsection*{Blocking the Precision Matrix}
The next step is to partition the precision matrix into a $2\times2$ block matrix. The matrix
$(\boldsymbol{Z}{'}\boldsymbol{V^{-1}Z})_{11}$ is the $T\times T$ submatrix, $(\boldsymbol{Z}^{'}\boldsymbol{V^{-1}Z})_{21}  =  (\boldsymbol{Z}^{'}\boldsymbol{V^{-1}Z})^{'}_{12}$ is the $T\times3$ submatrix and $(\boldsymbol{Z}{'}\boldsymbol{V^{-1}Z})_{22}$ is the $3\times3$ submatrix corresponding to the precision of the parameters of interest: $\hat{\theta}_1,\hat{\theta}_2,\text{ and } \hat{\theta}_3$. Using block matrix inversion \cite{TwoByTwoInverse}, we can find the inverse of this submatrix as $(\boldsymbol{Z}{'}\boldsymbol{V^{-1}Z})^{-1}_{22} = \left((\boldsymbol{Z}{'}\boldsymbol{V^{-1}Z})_{22} - (\boldsymbol{Z}{'}\boldsymbol{V^{-1}Z})_{21}(\boldsymbol{Z}{'}\boldsymbol{V^{-1}Z})_{11}^{-1}(\boldsymbol{Z}{'}\boldsymbol{V^{-1}Z})_{12}\right)^{-1}.$ We first obtain $(\boldsymbol{Z}{'}\boldsymbol{V^{-1}Z)})_{11}^{-1}$
using another variation of block matrix inversion \cite{TwoByTwoInverse} and Schur complements, yielding
\[(\boldsymbol{Z}{'}\boldsymbol{V^{-1}Z})_{11}^{-1} = \frac{1}{(f + gT)}
\begin{bmatrix}
\frac{(g + f)}{f} & -\boldsymbol{1{'}_{T-1}} \\
-\boldsymbol{1_{T-1}} &  \left(\boldsymbol{I_{T-1}} + \boldsymbol{1_{T-1}1{'}_{T-1}}\right)
\end{bmatrix}.
\]

Let $\boldsymbol{B} =\boldsymbol{I_{T-1}} + \boldsymbol{1_{T-1}1^{'}_{T-1}}$ and let $\boldsymbol{M}=(\boldsymbol{Z}{'}\boldsymbol{V^{-1}Z})_{21}(\boldsymbol{Z}{'}\boldsymbol{V^{-1}Z})_{11}^{-1}(\boldsymbol{Z}{'}\boldsymbol{V^{-1}Z})_{12}$. Using matrix multiplication, we can find 

\setlength{\abovedisplayskip}{2.5pt}
\setlength{\belowdisplayskip}{2.5pt}
\setlength{\abovedisplayshortskip}{2.5pt}
\setlength{\belowdisplayshortskip}{2.5pt}
\setlength\arraycolsep{2pt}
\medmuskip = 1mu
\[\hspace*{-1.2cm}\boldsymbol{M} = 
\frac{1}{(f+gT)} \begin{bmatrix}
y_1 &  \sum_{i=1}^{I}\frac{\boldsymbol{X^{'}_{i,-T}}}{\sigma_c^2}  - \sigma_{\alpha}^2h_1\boldsymbol{1^{'}_{T-1}}\\ y_2 &  \sum_{i=1}^{I}\frac{\boldsymbol{W^{'}_{i,-T}}}{\sigma_c^2}  - \sigma_{\alpha}^2h_2\boldsymbol{1^{'}_{T-1}} \\ y_3 &  \sum_{i=1}^{I}\frac{\boldsymbol{(XW)^{'}_{i,-T}}}{\sigma_c^2}  - \sigma_{\alpha}^2h_3\boldsymbol{1^{'}_{T-1}}
\end{bmatrix} \begin{bmatrix}
\frac{(g + f)}{f} & -\boldsymbol{1^{'}_{T-1}} \\
-\boldsymbol{1_{T-1}} &  \boldsymbol{B}
\end{bmatrix} \begin{bmatrix}
y_1 &  \sum_{i=1}^{I}\frac{\boldsymbol{X^{'}_{i,-T}}}{\sigma_c^2}  - \sigma_{\alpha}^2h_1\boldsymbol{1^{'}_{T-1}}\\ y_2 &  \sum_{i=1}^{I}\frac{\boldsymbol{W^{'}_{i,-T}}}{\sigma_c^2}  - \sigma_{\alpha}^2h_2\boldsymbol{1^{'}_{T-1}} \\ y_3 &  \sum_{i=1}^{I}\frac{\boldsymbol{(XW)^{'}_{i,-T}}}{\sigma_c^2}  - \sigma_{\alpha}^2h_3\boldsymbol{1^{'}_{T-1}}
\end{bmatrix}'\]

\setlength{\abovedisplayskip}{0pt}
\setlength{\belowdisplayskip}{0pt}
\setlength{\abovedisplayshortskip}{0pt}
\setlength{\belowdisplayshortskip}{0pt}
\footnotesize

\[\hspace{-1.5cm}=\frac{1}{(f+gT)} \begin{bmatrix}
y_1 &  \sum_{i=1}^{I}\frac{\boldsymbol{X^{'}_{i,-T}}}{\sigma_c^2}  - \sigma_{\alpha}^2h_1\boldsymbol{1^{'}_{T-1}}\\ y_2 &  \sum_{i=1}^{I}\frac{\boldsymbol{W^{'}_{i,-T}}}{\sigma_c^2}  - \sigma_{\alpha}^2h_2\boldsymbol{1^{'}_{T-1}}  \\ y_3 &  \sum_{i=1}^{I}\frac{\boldsymbol{(XW)^{'}_{i,-T}}}{\sigma_c^2}  - \sigma_{\alpha}^2h_3\boldsymbol{1^{'}_{T-1}}  
\end{bmatrix} 
\begin{bmatrix} \frac{g+f}{f}y_1 - \boldsymbol{1^{'}_{T-1}}\left(  \sum_{i=1}^{I}\frac{\boldsymbol{X_{i,-T}}}{\sigma_c^2}  - \sigma_{\alpha}^2h_1\boldsymbol{1_{T-1}}\right) & -y_1\boldsymbol{1_{T-1}} + \boldsymbol{B}\left( \sum_{i=1}^{I}\frac{\boldsymbol{X_{i,-T}}}{\sigma_c^2}  - \sigma_{\alpha}^2h_1\boldsymbol{1_{T-1}}\right) \\
\frac{g+f}{f}y_2 - \boldsymbol{1^{'}_{T-1}}\left(\sum_{i=1}^{I}\frac{\boldsymbol{W_{i,-T}}}{\sigma_c^2}  - \sigma_{\alpha}^2h_2\boldsymbol{1_{T-1}}\right) &
-y_2\boldsymbol{1_{T-1}} + \boldsymbol{B}\left( \sum_{i=1}^{I}\frac{\boldsymbol{W_{i,-T}}}{\sigma_c^2}  - \sigma_{\alpha}^2h_2\boldsymbol{1_{T-1}}\right) \\
\frac{g+f}{f}y_3 - \boldsymbol{1^{'}_{T-1}}\left(\sum_{i=1}^{I}\frac{\boldsymbol{(XW)_{i,-T}}}{\sigma_c^2}  - \sigma_{\alpha}^2h_3\boldsymbol{1_{T-1}}\right) &
-y_3\boldsymbol{1_{T-1}} + \boldsymbol{B}\left( \sum_{i=1}^{I}\frac{\boldsymbol{(XW)_{i,-T}}}{\sigma_c^2}  - \sigma_{\alpha}^2h_3\boldsymbol{1_{T-1}}\right)
\end{bmatrix}'.\]
\normalsize
\setlength{\abovedisplayskip}{2.5pt}
\setlength{\belowdisplayskip}{2.5pt}
\setlength{\abovedisplayshortskip}{2.5pt}
\setlength{\belowdisplayshortskip}{2.5pt}
Solving for the $-\boldsymbol{1_{T-1}'}\left(\sum_{i=1}^{I}\frac{\boldsymbol{X_{i,-T}}}{\sigma_c^2}  - \sigma_{\alpha}^2h_1\boldsymbol{1_{T-1}}\right)$ in the second matrix (similar for other elements), we obtain $$=\left(\sum_{i=j}^{T-1}\sum_{i=1}^{I}\frac{X_{ij}}{\sigma_c^2}  - h\sigma_{\alpha}^2 \boldsymbol{1_{T-1}}^{'}
\boldsymbol{1_{T-1}}\right) =\left(\sum_{i=j}^{T-1}\sum_{i=1}^{I}\frac{X_{ij}}{\sigma_c^2}  - \sigma_{\alpha}^2(T-1)\sum_{i=1}^I\sum_{j=1}^T\frac{X_{ij}}{\sigma_c^2(\sigma_c^2+T\sigma_{\alpha}^2)}\right)$$ $$=\left(\sum_{i=j}^{T}\sum_{i=1}^{I}\frac{\sigma_c^2X_{ij} + T\sigma_{\alpha}^2X_{ij}}{\sigma_c^2(\sigma_c^2+T\sigma_{\alpha}^2)} - \frac{\sigma_{\alpha}^2TX_{ij} - \sigma_{\alpha}^2X_{ij}}{\sigma_c^2(\sigma_c^2+T\sigma_{\alpha}^2)}\right)-\sum_{i=1}^{I}\frac{X_{iT}}{\sigma_c^2}=\left(\sum_{i=j}^{T}\sum_{i=1}^{I}\frac{X_{ij}}{(\sigma_c^2+T\sigma_{\alpha}^2)}\right) + \left(\sum_{i=j}^{T}\sum_{i=1}^{I}\frac{\sigma_{\alpha}^2X_{ij}}{\sigma_c^2(\sigma_c^2+T\sigma_{\alpha}^2)}\right) - \sum_{i=1}^{I}\frac{X_{iT}}{\sigma_c^2}$$$$=y_1 + \sigma_{\alpha}^2h_1 - \sum_{i=1}^{I}\frac{X_{iT}}{\sigma_c^2}.$$
Substituting this term back into $\boldsymbol{M}$ yields
$\boldsymbol{M} =\frac{1}{(f+gT)} \begin{bmatrix} m_{11} & m_{12} & m_{13}
\\m_{21} & m_{22} & m_{23}\\
m_{31} & m_{32} & m_{33}
\end{bmatrix}$ with elements
\footnotesize
\setlength\arraycolsep{2pt}
\medmuskip = 0.5mu 
$$\hspace{-1.2cm}m_{11} = y_1\left(\frac{g+f}{f}y_1 + \left(-y_1 - \sigma_{\alpha}^2h_1 + \sum_{i=1}^{I}\frac{X_{iT}}{\sigma_c^2}\right)\right) + \left( \sum_{i=1}^{I}\frac{\boldsymbol{X^{'}_{i,-T}}}{\sigma_c^2}  - \sigma_{\alpha}^2h_1\boldsymbol{1^{'}_{T-1}}\right)\left(-y_1\boldsymbol{1_{T-1}} + \boldsymbol{B}\left( \sum_{i=1}^{I}\frac{\boldsymbol{X_{i,-T}}}{\sigma_c^2}  - \sigma_{\alpha}^2h_1\boldsymbol{1_{T-1}}\right)\right) $$ $$\hspace{-1.2cm}m_{12} = y_1\left(\frac{g+f}{f}y_2 + \left(-y_2 - \sigma_{\alpha}^2h_2 + \sum_{i=1}^{I}\frac{W_{iT}}{\sigma_c^2}\right)\right) + \left(\sum_{i=1}^{I}\frac{\boldsymbol{X^{'}_{i,-T}}}{\sigma_c^2}  - \sigma_{\alpha}^2h_1\boldsymbol{1^{'}_{T-1}}\right)\left( -y_2\boldsymbol{1_{T-1}} + \boldsymbol{B}\left( \sum_{i=1}^{I}\frac{\boldsymbol{W_{i,-T}}}{\sigma_c^2}  - \sigma_{\alpha}^2h_2\boldsymbol{1_{T-1}}\right)\right) 
$$$$\hspace{-1.3cm}m_{13} = 
y_1\left(\frac{g+f}{f}y_3 + \left(-y_3 - \sigma_{\alpha}^2h_3 + \sum_{i=1}^{I}\frac{(XW)_{iT}}{\sigma_c^2}\right)\right) + \left(\sum_{i=1}^{I}\frac{\boldsymbol{X^{'}_{i,-T}}}{\sigma_c^2}  - \sigma_{\alpha}^2h_1\boldsymbol{1^{'}_{T-1}}\right)\left(-y_3\boldsymbol{1_{T-1}} + \boldsymbol{B}\left( \sum_{i=1}^{I}\frac{\boldsymbol{(XW)_{i,-T}}}{\sigma_c^2}  - \sigma_{\alpha}^2h_3\boldsymbol{1_{T-1}}\right)\right)
$$$$\hspace{-1.2cm}m_{21} =y_2\left(\frac{g+f}{f}y_1 - \left(y_1 + \sigma_{\alpha}^2h_1 - \sum_{i=1}^{I}\frac{X_{iT}}{\sigma_c^2}\right)\right) + \left(\sum_{i=1}^{I}\frac{\boldsymbol{W^{'}_{i,-T}}}{\sigma_c^2}  - \sigma_{\alpha}^2h_2\boldsymbol{1^{'}_{T-1}}\right)\left( -y_1\boldsymbol{1_{T-1}} + \boldsymbol{B}\left( \sum_{i=1}^{I}\frac{\boldsymbol{X_{i,-T}}}{\sigma_c^2}  - \sigma_{\alpha}^2h_1\boldsymbol{1_{T-1}}\right)\right)  $$
$$\hspace{-1.2cm}m_{22} = y_2\left(\frac{g+f}{f}y_2 - \left(y_2 + \sigma_{\alpha}^2h_2 - \sum_{i=1}^{I}\frac{W_{iT}}{\sigma_c^2}\right)\right) + \left(\sum_{i=1}^{I}\frac{\boldsymbol{W^{'}_{i,-T}}}{\sigma_c^2}  - \sigma_{\alpha}^2h_2\boldsymbol{1^{'}_{T-1}}\right)\left( -y_2\boldsymbol{1_{T-1}} + \boldsymbol{B}\left( \sum_{i=1}^{I}\frac{\boldsymbol{W_{i,-T}}}{\sigma_c^2}  - \sigma_{\alpha}^2h_2\boldsymbol{1_{T-1}}\right)\right)
$$$$\hspace{-1.2cm}m_{23} =y_2\left(\frac{g+f}{f}y_3 - \left(y_3 + \sigma_{\alpha}^2h_3 - \sum_{i=1}^{I}\frac{(XW)_{iT}}{\sigma_c^2}\right)\right) + \left(\sum_{i=1}^{I}\frac{\boldsymbol{W^{'}_{i,-T}}}{\sigma_c^2}  - \sigma_{\alpha}^2h_2\boldsymbol{1^{'}_{T-1}}\right)\left(-y_3\boldsymbol{1_{T-1}} + \boldsymbol{B}\left( \sum_{i=1}^{I}\frac{\boldsymbol{(XW)_{i,-T}}}{\sigma_c^2}  - \sigma_{\alpha}^2h_3\boldsymbol{1_{T-1}}\right)\right)
$$ $$\hspace{-1.2cm}m_{31} =
y_3\left(\frac{g+f}{f}y_1 - \left(y_1 + \sigma_{\alpha}^2h_1 - \sum_{i=1}^{I}\frac{X_{iT}}{\sigma_c^2}\right)\right) + \left(\sum_{i=1}^{I}\frac{\boldsymbol{XW^{'}_{i,-T}}}{\sigma_c^2}  - \sigma_{\alpha}^2h_3\boldsymbol{1^{'}_{T-1}}\right)\left( -y_1\boldsymbol{1_{T-1}} + \boldsymbol{B}\left( \sum_{i=1}^{I}\frac{\boldsymbol{X_{i,-T}}}{\sigma_c^2}  - \sigma_{\alpha}^2h_1\boldsymbol{1_{T-1}}\right)\right)  
$$ $$\hspace{-1.2cm}m_{32} = y_3\left(\frac{g+f}{f}y_2 - \left(y_2 + \sigma_{\alpha}^2h_2 - \sum_{i=1}^{I}\frac{W_{iT}}{\sigma_c^2}\right)\right) + \left(\sum_{i=1}^{I}\frac{\boldsymbol{XW^{'}_{i,-T}}}{\sigma_c^2}  - \sigma_{\alpha}^2h_3\boldsymbol{1^{'}_{T-1}}\right)\left( -y_2\boldsymbol{1_{T-1}} +\boldsymbol{B}\left( \sum_{i=1}^{I}\frac{\boldsymbol{W_{i,-T}}}{\sigma_c^2}  - \sigma_{\alpha}^2h_2\boldsymbol{1_{T-1}}\right)\right) $$ $$\hspace{-1.2cm}m_{33} = y_3\left(\frac{g+f}{f}y_3 - \left(y_3 + \sigma_{\alpha}^2h_3 - \sum_{i=1}^{I}\frac{(XW)_{iT}}{\sigma_c^2}\right)\right) + \left(\sum_{i=1}^{I}\frac{\boldsymbol{XW^{'}_{i,-T}}}{\sigma_c^2}  - \sigma_{\alpha}^2h_3\boldsymbol{1^{'}_{T-1}}\right)\left(-y_3\boldsymbol{1_{T-1}} + \boldsymbol{B}\left( \sum_{i=1}^{I}\frac{\boldsymbol{(XW)_{i,-T}}}{\sigma_c^2}  - \sigma_{\alpha}^2h_3\boldsymbol{1_{T-1}}\right)\right).$$

\subsection*{Simplifying the diagonal elements of $\boldsymbol{M}$}
\normalsize
The three diagonal elements of $\boldsymbol{M}$ have the same form, so we will solve for the (1,1) element of the matrix corresponding to treatment 1 and apply the form to all three diagonal elements. Let $\eta_1 = y_1 + \sigma_{\alpha}^2h_1 - \sum_{i=1}^{I}\frac{X_{iT}}{\sigma_c^2}$, then the diagonal element is:
\setlength{\abovedisplayskip}{4pt}
\setlength{\belowdisplayskip}{4pt}
\setlength{\abovedisplayshortskip}{4pt}
\setlength{\belowdisplayshortskip}{4pt}
\begin{flalign}=\frac{1}{(f+gT)}\Bigg( y_1\left(\frac{g+f}{f}y_1 - \eta_1\right) + \left( \sum_{i=1}^{I}\frac{\boldsymbol{X_{i,-T}}}{\sigma_c^2}  - \sigma_{\alpha}^2h_1\boldsymbol{1_{T-1}}\right)\left(-y_1\boldsymbol{1_{T-1}} + \boldsymbol{B}\left( \sum_{i=1}^{I}\frac{\boldsymbol{X_{i,-T}}}{\sigma_c^2}  - \sigma_{\alpha}^2h_1\boldsymbol{1_{T-1}}\right)\right)\Bigg)\nonumber\end{flalign}
$$=\frac{1}{(f+gT)}\left( \frac{y_1^2(f+g)}{f} - 2y_1\eta_1 +  \left(\sum_{i=1}^{I}\frac{\boldsymbol{X_{i,-T}^{'}}}{\sigma_c^2}  - \sigma_{\alpha}^2h_1\boldsymbol{1_{T-1}^{'}}\right)\boldsymbol{B}\left(\sum_{i=1}^{I}\frac{\boldsymbol{X_{i,-T}}}{\sigma_c^2}  - \sigma_{\alpha}^2h_1\boldsymbol{1_{T-1}}\right)\right).
$$
\setlength{\abovedisplayskip}{4pt}
\setlength{\belowdisplayskip}{4pt}
\setlength{\abovedisplayshortskip}{4pt}
\setlength{\belowdisplayshortskip}{4pt}
Further, 
\[\left(\sum_{i=1}^{I}\frac{\boldsymbol{X_{i,-T}^{'}}}{\sigma_c^2}  - \frac{l_1 - y_1}{T}\boldsymbol{1_{T-1}^{'}}\right)\boldsymbol{B}\left(\sum_{i=1}^{I}\frac{\boldsymbol{X_{i,-T}}}{\sigma_c^2}  - \frac{l_1 - y_1}{T}\boldsymbol{1_{T-1}}\right)\]\[=\left(\sum_{i=1}^{I}\frac{\boldsymbol{X_{i,-T}^{'}}}{\sigma_c^2}  - \frac{l_1}{T}\boldsymbol{1_{T-1}^{'}}\right)\boldsymbol{B}\left(\sum_{i=1}^{I}\frac{\boldsymbol{X_{i,-T}}}{\sigma_c^2}  - \frac{l_1}{T}\boldsymbol{1_{T-1}}\right) + \frac{y_1}{T}\boldsymbol{1_{T-1}^{'}}\boldsymbol{B}\frac{y_1}{T}\boldsymbol{1_{T-1}} + 2 \frac{y_1}{T}\boldsymbol{1_{T-1}^{'}}\boldsymbol{B}\left(\sum_{i=1}^{I}\frac{\boldsymbol{X_{i,-T}}}{\sigma_c^2}  - \frac{l_1}{T}\boldsymbol{1_{T-1}}\right).\]
\normalsize

We solve for each term individually. We will make use of the relationship $\sum_{i=1}^{I}\sum_{j=1}^{T-1}\frac{X_{ij}}{\sigma_c^2} = l_1- \sum_{i=1}^{I}\frac{X_{iT}}{\sigma_c^2}$. Solving for the first term, $\left(\sum_{i=1}^{I}\frac{\boldsymbol{X_{i,-T}^{'}}}{\sigma_c^2}  - \frac{l_1}{T}\boldsymbol{1_{T-1}^{'}}\right)\boldsymbol{B}\left(\sum_{i=1}^{I}\frac{\boldsymbol{X_{i,-T}}}{\sigma_c^2}  - \frac{l_1}{T}\boldsymbol{1_{T-1}}\right)$ 
\setlength{\abovedisplayskip}{2.5pt}
\setlength{\belowdisplayskip}{2.5pt}
\setlength{\abovedisplayshortskip}{2.5pt}
\setlength{\belowdisplayshortskip}{2.5pt}
\[=\sum_{j=1}^{T-1}\left(\sum_{i=1}^{I}\frac{X_{ij}}{\sigma_c^2}\right)^2 + l_1^2 - 2l_1\sum_{i=1}^{I}\frac{X_{iT}}{\sigma_c^2} + \left(\sum_{i=1}^{I}\frac{X_{iT}}{\sigma_c^2}\right)^2 - 2l_1^2 +
2l_1\sum_{i=1}^{I}\frac{X_{ij}}{\sigma_c^2} + l_1^2 - \frac{l_1^2}{T} \;\;\; = \;\;\; \sum_{j=1}^{T}\left(\sum_{i=1}^{I}\frac{X_{ij}}{\sigma_c^2}\right)^2- \frac{l_1^2}{T}. \]

Letting $w_1 = \sum_{j=1}^{T}\left(\sum_{i=1}^{I}\frac{X_{ij}}{\sigma_c^2}\right)^2$, the first term becomes $w_1 - \frac{l_1^2}{T}$. Similarly for other diagonal elements, let  $w_2 = \sum_{j=1}^{T}\left(\sum_{i=1}^{I}\frac{W_{ij}}{\sigma_c^2}\right)^2$ and  $w_3 = \sum_{j=1}^{T}\left(\sum_{i=1}^{I}\frac{(XW)_{ij}}{\sigma_c^2}\right)^2$. Solving the second term, $\frac{y_1}{T}\boldsymbol{1_{T-1}^{'}}\boldsymbol{B}\frac{y_1}{T}\boldsymbol{1_{T-1}}$, yields
\[=\frac{y_1^2}{T^2}\boldsymbol{1_{T-1}^{'}}(\boldsymbol{I_{T-1}} + \boldsymbol{1_{T-1}1{'}_{T-1}})\boldsymbol{1_{T-1}}\;=\;\frac{y_1^2}{T^2}\big((T-1) + (T-1)\boldsymbol{1_{T-1}^{'}}\boldsymbol{1_{T-1}}\big) \;\; = \;\; \frac{y_1^2}{T^2}(T-1 + T^2 -2T + 1) \;= \; \frac{y_1^2}{T}(T - 1).\]

Solving for the third term, $\frac{2y_1}{T}\boldsymbol{1_{T-1}^{'}}\boldsymbol{B}\left(\sum_{i=1}^{I}\frac{\boldsymbol{X_{i,-T}}}{\sigma_c^2}  - \frac{l_1}{T}\boldsymbol{1_{T-1}}\right)$, yields
\setlength{\abovedisplayshortskip}{5pt}
\setlength{\belowdisplayshortskip}{5pt}
\setlength{\abovedisplayskip}{5pt}
\setlength{\belowdisplayskip}{5pt}
$$\frac{2y_1}{T}(\boldsymbol{1_{T-1}^{'}} + (T-1)\boldsymbol{1_{T-1}^{'}})\left(\sum_{i=1}^{I}\frac{\boldsymbol{X_{i,-T}}}{\sigma_c^2}  - \frac{l_1}{T}\boldsymbol{1_{T-1}}\right)\;=\;2y_1\boldsymbol{1_{T-1}^{'}}\left(\sum_{i=1}^{I}\frac{\boldsymbol{X_{i,-T}}}{\sigma_c^2}  - \frac{l_1}{T}\boldsymbol{1_{T-1}}\right)$$ $$=2y_1\left(-\sum_{i=1}^{I}\frac{X_{iT}}{\sigma_c^2} - \frac{l_1}{T} + \frac{y_1}{T} - \frac{y_1}{T}\right) = 2y_1\left(\frac{y_1}{T} + \sigma_{\alpha}^2h_1 - \sum_{i=1}^{I}\frac{X_{iT}}{\sigma_c^2} \right).$$

Substituting these terms back in the diagonal element of $\boldsymbol{M}$, we obtain
\begin{align}
=\frac{1}{(f+gT)}\Bigg( \frac{y_1^2(f+g)}{f} - 2y_1\eta_1 + \nonumber \left(\sum_{i=1}^{I}\frac{\boldsymbol{X_{i,-T}^{'}}}{\sigma_c^2}  -  \sigma_{\alpha}^2h_1\boldsymbol{1_{T-1}^{'}}\right)\boldsymbol{B}\left(\sum_{i=1}^{I}\frac{\boldsymbol{X_{i,-T}}}{\sigma_c^2}  - \sigma_{\alpha}^2h_1\boldsymbol{1_{T-1}}\right)\Bigg)\nonumber\end{align}
\begin{align}= \frac{1}{(f+gT)}\left(\frac{y_1^2(f+g)}{f} - 2y_1\left(y_1 + \sigma_{\alpha}^2h_1 + \sum_{i=1}^{I}\frac{X_{iT}}{\sigma_c^2}\right) +  w_1-\frac{l_1^2}{T} + 2y_1\left(\frac{y_1}{T} + \sigma_{\alpha}^2h_1 - \sum_{i=1}^{I}\frac{X_{iT}}{\sigma_c^2}\right) + \frac{y_1^2}{T}(T-1)\right)\nonumber\end{align} \begin{align}= \frac{1}{(f+gT)}\left(y_1^2\left(\frac{f+gT}{fT}\right)  +  w_1-\frac{l_1^2}{T} \right)\;= \;\frac{y_1^2}{fT}  +  \frac{1}{(f+gT)}\left(w_1-\frac{l_1^2}{T}\right).\nonumber\end{align}

The second and third diagonal elements of the matrix $\boldsymbol{M}$ will have the same form.

\subsection*{Simplifying the off-diagonal elements of $\boldsymbol{M}$}
To solve for the off-diagonal values, we consider one value, $m_{21}$, and apply the results to the other off-diagonals. The value of $m_{21}$ is 

\footnotesize
\setlength{\arraycolsep}{2.5pt}
\medmuskip = 2mu 
$$ \frac{1}{(f+gT)} \left(y_2\left(\frac{g+f}{f}y_1 - y_1 - \sigma_{\alpha}^2h_1 + \sum_{i=1}^{I}\frac{X_{iT}}{\sigma_c^2}\right) + \left(\sum_{i=1}^{I}\frac{\boldsymbol{W_{i,-T}}^{'}}{\sigma_c^2}  - \sigma_{\alpha}^2h_2\boldsymbol{1_{T-1}^{'}}\right)\left( -y_1\boldsymbol{1_{T-1}} +\boldsymbol{B}\left( \sum_{i=1}^{I}\frac{\boldsymbol{X_{i,-T}}}{\sigma_c^2}  - \sigma_{\alpha}^2h_1\boldsymbol{1_{T-1}}\right)\right)\right).$$

\setlength{\arraycolsep}{3pt}
\medmuskip = 4mu
\normalsize
We will first simplify the term $y_2\left(\frac{g+f}{f}y_1 -y_1 - \sigma_{\alpha}^2h_1 + \sum_{i=1}^{I}\frac{X_{iT}}{\sigma_c^2}\right) \;\;= \;\;  
y_1y_2\frac{f+g}{f} - y_1y_2 - \sigma_{\alpha}^2h_1y_2 + y_2\sum_{i=1}^{I}\frac{X_{iT}}{\sigma_c^2}.$ To simplify the second term,
$\left(\sum_{i=1}^{I}\frac{\boldsymbol{W_{i,-T}}^{'}}{\sigma_c^2}  - \sigma_{\alpha}^2h_2\boldsymbol{1_{T-1}^{'}}\right)\left( -y_1\boldsymbol{1_{T-1}} + \boldsymbol{B}\left( \sum_{i=1}^{I}\frac{\boldsymbol{X_{i,-T}}}{\sigma_c^2}  - \sigma_{\alpha}^2h_1\boldsymbol{1_{T-1}}\right)\right)$, we will break up this multiplication into different terms. The first step is multiplying two terms, yielding  $$\left(\sum_{i=1}^{I}\frac{\boldsymbol{W_{i,-T}^{'}}}{\sigma_c^2}  - \sigma_{\alpha}^2h_2\boldsymbol{1_{T-1}^{'}}\right)(-y_1\boldsymbol{1_{T-1}}) \;=\; -y_1\sum_{j=1}^{T-1}\sum_{i=1}^{I}\frac{W_{ij}}{\sigma_c^2}  + (T-1) \sigma_{\alpha}^2h_2y_1.$$

The rest of the multiplication $\left(\sum_{i=1}^{I}\frac{\boldsymbol{W_{i,-T}}^{'}}{\sigma_c^2}  - \sigma_{\alpha}^2h_2\boldsymbol{1_{T-1}^{'}}\right)\left(\boldsymbol{B}\left( \sum_{i=1}^{I}\frac{\boldsymbol{X_{i,-T}}}{\sigma_c^2}  - \sigma_{\alpha}^2h_1\boldsymbol{1_{T-1}}\right)\right)$ simplifies to
$$=\left(\sum_{i=1}^{I}\frac{\boldsymbol{W_{i,-T}}^{'}}{\sigma_c^2}  - \frac{l_2 - y_2}{T}\boldsymbol{1_{T-1}^{'}}\right)\left(\left(\boldsymbol{I_{T-1}} + \boldsymbol{1_{T-1}1_{T-1}^{'}}\right)\left( \sum_{i=1}^{I}\frac{\boldsymbol{X_{i,-T}}}{\sigma_c^2}  - \frac{l_1 - y_1}{T}\boldsymbol{1_{T-1}}\right)\right)$$
\setlength{\abovedisplayshortskip}{0pt}
\setlength{\belowdisplayshortskip}{0pt}
\setlength{\abovedisplayskip}{4pt}
\setlength{\belowdisplayskip}{2pt}
$$=\left(\sum_{i=1}^{I}\frac{\boldsymbol{W_{i,-T}}^{'}}{\sigma_c^2}  - \frac{l_2}{T}\boldsymbol{1_{T-1}^{'}} + \frac{y_2}{T}\boldsymbol{1_{T-1}^{'}}\right)\left(\left(\boldsymbol{I_{T-1}} + \boldsymbol{1_{T-1}1_{T-1}^{'}}\right)\left( \sum_{i=1}^{I}\frac{\boldsymbol{X_{i,-T}}}{\sigma_c^2}  - \frac{l_1}{T}\boldsymbol{1_{T-1}} + \frac{y_1}{T}\boldsymbol{1_{T-1}}\right)\right)$$
\begin{align*} =\left(\sum_{i=1}^{I}\frac{\boldsymbol{W_{i,-T}}^{'}}{\sigma_c^2}  - \frac{l_2}{T}\boldsymbol{1_{T-1}}^{'}\right)\left(\left(\boldsymbol{I_{T-1}} + \boldsymbol{1_{T-1}1_{T-1}^{'}}\right)\left( \sum_{i=1}^{I}\frac{\boldsymbol{X_{i,-T}}}{\sigma_c^2}  - \frac{l_1}{T}\boldsymbol{1_{T-1}}\right)\right) + \\ \frac{y_2}{T}\boldsymbol{1_{T-1}^{'}}(\boldsymbol{I_{T-1}} + \boldsymbol{1_{T-1}1_{T-1}^{'}})\frac{y_1}{T}\boldsymbol{1_{T-1}} + \frac{y_2}{T}\boldsymbol{1_{T-1}^{'}}(\boldsymbol{I_{T-1}} + \boldsymbol{1_{T-1}1_{T-1}^{'}})\left(\sum_{i=1}^{I}\frac{\boldsymbol{X_{i,-T}}}{\sigma_c^2}  -\frac{l_1}{T}\boldsymbol{1_{T-1}}\right) + \\ \left(\sum_{i=1}^{I}\frac{\boldsymbol{W_{i,-T}}^{'}}{\sigma_c^2}  - \frac{l_1}{T}\boldsymbol{1_{T-1}^{'}}\right)(\boldsymbol{I_{T-1}}+\boldsymbol{1_{T-1}1_{T-1}^{'}})(\frac{y_1}{T}\boldsymbol{1_{T-1}}).
\end{align*}
We break this multiplication up by individual terms, with the first term simplifying to
\setlength{\abovedisplayshortskip}{5pt}
\setlength{\belowdisplayshortskip}{5pt}
\setlength{\abovedisplayskip}{5pt}
\setlength{\belowdisplayskip}{5pt}
$$\left(\sum_{i=1}^{I}\frac{\boldsymbol{W_{i,-T}}^{'}}{\sigma_c^2}  - \frac{l_2}{T}\boldsymbol{1_{T-1}^{'}}\right)\left(\boldsymbol{B}\left( \sum_{i=1}^{I}\frac{\boldsymbol{X_{i,-T}}}{\sigma_c^2}  - \frac{l_1}{T}\boldsymbol{1_{T-1}}\right)\right)$$ \begin{align*}
     = \sum_{j=1}^{T-1}\left(\sum_{i=1}^{I}\frac{W_{ij}}{\sigma_c^2}\sum_{i=1}^{I}\frac{X_{ij}}{\sigma_c^2}\right) + \left(\sum_{j=1}^{T-1}\sum_{i=1}^{I}\frac{W_{ij}}{\sigma_c^2}\right)\left(\sum_{j=1}^{T-1}\sum_{i=1}^{I}\frac{X_{ij}}{\sigma_c^2}\right) -  \frac{l_1}{T}\sum_{j=1}^{T-1}\sum_{i=1}^{I}\frac{W_{ij}}{\sigma_c^2} - \\ \frac{l_1(T-1)}{T}\sum_{j=1}^{T-1}\sum_{i=1}^{I}\frac{W_{ij}}{\sigma_c^2} -
\frac{l_2}{T}\sum_{j=1}^{T-1}\sum_{i=1}^{I}\frac{X_{ij}}{\sigma_c^2} - \frac{l_2(T-1)}{T}\sum_{j=1}^{T-1}\sum_{i=1}^{I}\frac{X_{ij}}{\sigma_c^2} + \frac{l_1l_2(T-1)}{T^2} + \frac{l_1l_2(T-1)^2}{T^2}\end{align*}
$$\hspace{-0.5cm}=\sum_{j=1}^{T-1}\left(\sum_{i=1}^{I}\frac{W_{ij}}{\sigma_c^2}\sum_{i=1}^{I}\frac{X_{ij}}{\sigma_c^2}\right) + \left(\sum_{j=1}^{T-1}\sum_{i=1}^{I}\frac{W_{ij}}{\sigma_c^2}\right)\left(\sum_{j=1}^{T-1}\sum_{i=1}^{I}\frac{X_{ij}}{\sigma_c^2}\right) - l_1\sum_{j=1}^{T-1}\sum_{i=1}^{I}\frac{W_{ij}}{\sigma_c^2} - l_2\sum_{j=1}^{T-1}\sum_{i=1}^{I}\frac{X_{ij}}{\sigma_c^2} +\frac{(T-1)l_1l_2}{T}$$
$$ =w_{XW} - \sum_{j=1}^{T}\left(\sum_{i=1}^{I}\frac{W_{ij}}{\sigma_c^2}\sum_{i=1}^{I}\frac{X_{ij}}{\sigma_c^2}\right)  - \frac{l_1l_2}{T}, \text{ where } \;\;w_{XW} = \sum_{j=1}^{T}\left(\sum_{i=1}^{I}\frac{W_{ij}}{\sigma_c^2}\sum_{i=1}^{I}\frac{X_{ij}}{\sigma_c^2}\right).$$

Let $w_{X(XW)} = \sum_{j=1}^{T}\left(\sum_{i=1}^{I}\frac{X_{ij}}{\sigma_c^2}\sum_{i=1}^{I}\frac{(XW)_{ij}}{\sigma_c^2}\right)$ and   $w_{W(XW)} = \sum_{j=1}^{T}\left(\sum_{i=1}^{I}\frac{W_{ij}}{\sigma_c^2}\sum_{i=1}^{I}\frac{(XW)_{ij}}{\sigma_c^2}\right)$ for the other off-diagonal elements. Simplifying the remaining terms, $$\frac{y_2}{T}\boldsymbol{1_{T-1}^{'}}\boldsymbol{B}\frac{y_1}{T}\boldsymbol{1_{T-1}} + \frac{y_2}{T}\boldsymbol{1_{T-1}^{'}}\boldsymbol{B}\left(\sum_{i=1}^{I}\frac{\boldsymbol{X_{i,-T}}}{\sigma_c^2}  -\frac{l_1}{T}\boldsymbol{1_{T-1}}\right) +  \left(\sum_{i=1}^{I}\frac{\boldsymbol{W_{i,-T}}^{'}}{\sigma_c^2}  - \frac{l_2}{T}\boldsymbol{1_{T-1}^{'}}\right)\boldsymbol{B}(\frac{y_1}{T}\boldsymbol{1_{T-1}})$$ 
$$=\frac{y_1y_2(T-1)}{T} + y_2\left(\frac{y_1}{T} + \sigma_{\alpha}^2h_1 - \sum_{i=1}^{I}\frac{X_{iT}}{\sigma_c^2}\right) + y_1\left(\frac{y_2}{T} + \sigma_{\alpha}^2h_2 - \sum_{i=1}^{I}\frac{Y_{iT}}{\sigma_c^2}\right).$$
Putting all the terms together for the off-diagonals, we get:
\begin{align*}
= y_1y_2\frac{f+g}{f} - y_1y_2 - \sigma_{\alpha}^2h_1y_2 + y_2\sum_{i=1}^{I}\frac{X_{iT}}{\sigma_c^2} - y_1\sum_{j=1}^{T-1}\sum_{i=1}^{I}\frac{W_{ij}}{\sigma_c^2}  + (T-1) \sigma_{\alpha}^2h_2y_1 + \\ \frac{y_1y_2(T-1)}{T} + y_2\left(\frac{y_1}{T} +\sigma_{\alpha}^2h_1 - \sum_{i=1}^{I}\frac{X_{iT}}{\sigma_c^2}\right) + y_1\left(\frac{y_2}{T} + \sigma_{\alpha}^2h_2 - \sum_{i=1}^{I}\frac{Y_{iT}}{\sigma_c^2}\right) + w_{xy} - \frac{l_1l_2}{T}
\end{align*}
\begin{align*} = y_1y_2\frac{f+gT+fT}{fT}  - y_1l_2  + T\sigma_{\alpha}^2h_2y_1  + w_{xy} - \frac{l_1l_2}{T}
\; = \; y_1y_2\left(\frac{f+gT}{fT}\right)  + w_{xy} - \frac{l_1l_2}{T}.
\end{align*}	
\normalsize
Multiplying this by the constant $\frac{1}{f+gT}$, the off-diagonal element becomes $\frac{y_1y_2}{fT}  + \frac{1}{f+gT}\left(w_{xy} - \frac{l_1l_2}{T}\right).$

\subsection*{Final Algebra}
We now have a simplified expression for $\boldsymbol{M}$. To obtain $(\boldsymbol{Z}{'}\boldsymbol{V^{-1}Z})_{22}^{-1}$, we calculate $(\boldsymbol{Z}{'}\boldsymbol{V^{-1}Z})_{22} - \boldsymbol{M}$ and take the inverse of this matrix. We have $(\boldsymbol{Z}{'}\boldsymbol{V^{-1}Z})_{22}-\boldsymbol{M}$ equal to \[= 
\begin{bmatrix} l_1 - z_1 - \frac{y_1^2}{fT} - \frac{1}{f+gT}\left(w_1 -\frac{l_1^2}{T}\right) & q_1 - 
\frac{y_1y_2}{fT}  - \frac{1}{f+gT}\left(w_{XW} - \frac{l_1l_2}{T}\right)
 
& q_2 - 
\frac{y_1y_3}{fT}  - \frac{1}{f+gT}\left(w_{X(XW)} - \frac{l_1l_3}{T}\right)
\\
q_1 - \frac{y_1y_2}{fT} - \frac{1}{f+gT}\left(w_{XW} \frac{l_1l_2}{T}\right) &  l_2 - z_2 - \frac{y_2^2}{fT} - \frac{1}{f+gT}\left(w_2 -\frac{l_2^2}{T}\right) & q_3 - 
\frac{y_2y_3}{fT}  - \frac{1}{f+gT}\left(w_{W(XW)} - \frac{l_2l_3}{T}\right)\\
q_2 - \frac{y_1y_3}{fT} - \frac{1}{f+gT}\left(w_{X(XW)} \frac{l_1l_3}{T}\right) & q_3 - \frac{y_2y_3}{fT} - \frac{1}{f+gT}\left(w_{W(XW)} \frac{l_2l_3}{T}\right)& l_3 - z_3 - \frac{y_3^2}{fT} - \frac{1}{f+gT}\left(w_3 -\frac{l_3^2}{T}\right)
\end{bmatrix}.
\]

We can now solve for $(\boldsymbol{Z}{'}\boldsymbol{V^{-1}Z})_{22}^{-1} $ by taking the inverse of this $3\times3$ matrix. If treatment effects are assumed to be additive and an interaction term is not included, we can solve for the upper $2\times2$ matrix for the variance-covariance matrix of the regression coefficients for two main treatment effects. For a one treatment repeated cross-sectional stepped wedge design, the variance of the treatment effect would be the reciprocal of the diagonal term, $ l_1 - z_1 - \frac{y_1^2}{fT} - \frac{1}{f+gT}\left(w_1 -\frac{l_1^2}{T}\right),$ which can be shown to be the same variance as solved for by Hussey and Hughes \cite{HusseyHughes}.

\section*{Conflict of Interest}
No potential conflict of interest was reported by the authors.

\bibliography{template}

\end{document}